\documentclass[a4paper,titlepage]{article}
\usepackage{mitpress}
\usepackage{amsmath}
\usepackage{graphicx}
\usepackage{amsfonts}
\usepackage{amssymb}

\newdimen\dummy
\dummy=\oddsidemargin
\begin{document}

\title{Space-Time Approach to Scattering from Many Body Systems}
\author{R. G\"{a}hler$^{\ast}$, J. Felber$^{\ast}$, F. Mezei$^{\#}$ and R. Golub$^{\#}$\\$^{\#}$Hahn Meitner Institut, Glienickerstr. 100, D-14109 Berlin\\$^{\ast}$TU-M\"{u}nchen, Fakult\"{a}t f\"{u}r Physik, D-85747 Garching}
\maketitle
\begin{abstract}
We present scattering from many body systems in a new light. In place of the
usual van Hove treatment, (applicable to a wide range of scattering processes
using both photons and massive particles) based on plane waves, we calculate
the scattering amplitude as a space-time integral over the scattering sample
for an incident wave characterized by its correlation function which results
from the shaping of the wave field by the apparatus. Instrument resolution
effects - seen as due to the loss of correlation caused by the path
differences in the different arms of the instrument are automatically included
and analytic forms of the resolution function for different instruments are
obtained. Each element of the apparatus is associated with a correlation
length (or time). These correlation lengths, determined by the dimensions of
the apparatus are generally much smaller than these dimensions and larger than
the wavelength. As is well known, these are the conditions for the validity of
geometrical optics so that the conventional treatment, where the scattering is
calculated by the van Hove plane wave approach and the trajectories through
the instrument are treated classically, is usually valid. In the present
approach analytic expressions for the correlation functions are obtained. The
intersection of the moving correlation volumes (those regions where the
correlation functions are significant) associated with the different elements
of the apparatus determines the maximum correlation lengths (times) that can
be observed in a sample, and hence, the momentum (energy) resolution of the
measurement. This geometrical picture of moving correlation volumes derived by
our technique shows how the interaction of the scatterer with the wave field
shaped by the apparatus proceeds in space and time. Matching of the
correlation volumes so as to maximize the intersection region yields a
transparent, graphical method of instrument design. PACS: 03.65.Nk, 3.80 +r,
03.75, 61.12.B
\end{abstract}

\section{Introduction}

The usual treatment of scattering from many particle systems assumes the
initial beam to be represented by a plane wave and shows that the scattering
cross section is proportional to the Fourier transform of the correlation
function of the density fluctuations. For inelastic (energy changing)
scattering this is the van Hove time dependent correlation function, in the
case of elastic scattering it is the static pair distribution function
\cite{vanh}. The usual treatment then integrates the cross section for
monochromatic plane waves over the spread of momentum in the incident and
scattered beams.

In this work we show that another approach is both possible and interesting.
As we do the entire calculation in real space-time our approach sheds new
light on the emergence of the Fourier transform in the scattering cross
section. A crucial feature of our approach is that the influence of beam
preparation on the resolution of the measurement is expressed by the
correlation volume of the incident beam, which in turn, is determined by the
optical (in the sense of normal \cite{BW} and time dependent \cite{ggtdo}
optics) properties of the beam preparation elements, \emph{e.g. }slits and choppers.

We will see that the lateral width of the beam correlation volume determines
the possible spatial extent of the correlations which can be observed in a
given experiment, and is inversely proportional to the $q$ resolution, as in
the conventional treatment. The longitudinal extent of the beam correlation
volume or the correlation time of the incident beam likewise sets the limit to
the system correlation times which can be observed. The correlation volume
$V_{c}$ is the conjugate volume to the momentum space volume $V_{p}$ and
$V_{p}V_{c}\approx\hbar^{3}$.

In a previous work \cite{nsplit} we have discussed the neutron spin echo
spectrometer \cite{mezse} from a similar point of view. We have shown that in
that case the equivalent correlation volume is split longitudinally by the
'spin echo time', $\tau_{\text{NSE}}$, which can be much longer than the
correlation time of the incident beam. This splitting of the correlation
volume results in the fact that the spin echo spectrometer measures the time
dependence of the scattering system correlation function directly. When the
incoming beam correlation volume is continuous (not split), then the scattered
wave is an integral over all space-time separations falling within the
correlation volume of the incident beam and the result is the Fourier
transform of the space-time correlation function of the scattering system as
will be seen below.

\subsection{Correlation functions and scattering}

\smallskip The concept of correlation (coherence) volume has been applied to
visible light optics for more than 50 years including such problems as the
effect of the illumination on the resolution of a microscope . The correlation
properties of a beam as a result of the optical properties of its preparing
devices are known as the van Cittert-Zernike theorem \cite{BW}, \cite{champ}
and \cite{cz} and we will see how this can be applied to both time -dependent
and -independent scattering.

A frequently used technique for deconvoluting the instrument resolution from
the measured data is to use the Fourier transform of the resolution function
\cite{pynnco}. In the present work we show the physical significance of that
Fourier transform and show how it is determined by the optical properties of
the beam. It has been shown in \cite{pynnco} that this Fourier transform may
contain more information than \emph{e.g. }the width of the resolution
function. In fact it is the correlation function at long distances and times
that is often of the greatest importance. An additional feature of our
treatment is that the instrumental resolution is seen as an integral part of
the experiment not something that is put in \emph{ad hoc }after the calculation.

It is well known that scattering experiments depend only on the spectrum
$\left|  A\left(  \overrightarrow{k}\right)  \right|  ^{2}$ of the incident
beam and hence can be considered as the incoherent superposition of the
results of experiments done with plane waves distributed according to this
spectrum. This is easily seen in the context of the present work where we show
that the scattering cross section depends on the auto-correlation function of
the incident beam. The Fourier transform of this auto-correlation is $\left|
A\left(  \overrightarrow{k}\right)  \right|  ^{2}$. In addition it is usually
assumed that sources produce 'chaotic' beams, \emph{i.e.} beams with
$\left\langle A\left(  \overrightarrow{k_{1}}\right)  A^{*}\left(
\overrightarrow{k_{2}}\right)  \right\rangle =C\delta\left(  \overrightarrow
{k_{1}}-\overrightarrow{k_{2}}\right)  $, \cite{ns}, so that in any case it is
only $\left|  A\left(  \overrightarrow{k}\right)  \right|  ^{2}$ which is
significant. This is the case for 'stationary' beams where $\left\langle
\left|  \psi\left(  x\right)  \right|  ^{2}\right\rangle $ is independent of
$x$.

These points, and the possibilities of measuring $\left\langle A\left(
\overrightarrow{k_{1}}\right)  A^{*}\left(  \overrightarrow{k_{2}}\right)
\right\rangle $ have been discussed in \cite{cohbab} and the references
therein, and have been emphasized by Mezei \cite{mezhouch}. For these reasons
we feel that it is more convenient to talk about correlation volumes rather
than coherence volumes as has been the practice in optics \cite{BW} since it
seems reasonable to reserve the term coherence for those cases where
$\left\langle A\left(  \overrightarrow{k_{1}}\right)  A^{*}\left(
\overrightarrow{k_{2}}\right)  \right\rangle $ plays an important role.

The correlation length $\delta_{c}$ discussed here and in section
\ref{elscatt} has real physical meaning (as of course does its equivalent the
spectral width $\Delta k$). This can be seen by considering that in order to
observe the famous two slit interference pattern (Young) it is necessary to
illuminate the two slits with correlated radiation. This is normally produced
by having the two slits located in the main peak of the diffraction pattern
produced by a third slit up-stream from the pair producing the interference
pattern. The van Cittert- Zernike theorem is the rigorous expression of this
point. The two point correlation function can be measured with such a double
slit experiment by measuring the modulation (visibility) of the interference
pattern as a function of the slit separation. Of course the physical situation
can be described according to the spectral viewpoint, where the incident beam
is considered as an incoherent superposition of different $\overrightarrow{k}$
vectors, by considering that for increasing slit separations the interference
patterns for the different incident $\overrightarrow{k}$ vectors begin to
cancel out.

Recent experiments involving M\"{o}ssbauer scattering of synchrotron radiation
have been described in terms of the beam correlation volume \cite{synch}.
These experiments can also be interpreted in terms of $\left|  A\left(
\overrightarrow{k}\right)  \right|  ^{2}$. However the authors have shown that
the concept of a coherence (correlation) length is very useful for
understanding this exciting experiment.

In the following we shall see that these ideas have a rigorous basis. We have
presented the main features of these ideas with some physical examples in
\cite{interlak}.

\subsection{Examples of correlation lengths and times}

We now discuss two simple examples to show the magnitudes of the correlation
lengths (times) introduced by the shaping of the wave fields produced by the
optical elements of the apparatus. As formulated by Marathay ''..the field
from a non-coherent (\emph{uncorrelated) }source acquires coherence (\emph{a
finite correlation length) }by the very process of propagation.'' (see
\cite{marat}, p.105.) This is nothing else than what occurs when a random
'white noise' signal acquires a finite correlation time as a result of passing
through a low pass filter, something which can easily be observed with an oscilloscope.

For elastic scattering, a beam collimated with slits of width $a$ separated by
a distance $L$, will have an angular width of $\Delta\theta\sim a/L$ and a
spread in momentum perpendicular to the beam of $\Delta k_{y}\sim k_{x}%
\Delta\theta\sim k_{x}a/L$. This will result in an uncertainty in momentum
transfer $q$ of $\Delta q\sim\Delta k_{y}$.

On the other hand, seen according to quantum mechanics, a beam going through a
slit of width $a$ will have an uncertainty in momentum $\Delta k_{y}^{\prime
}\sim1/a$ and this will result in a lateral spreading over a distance
$\delta\sim\Delta k_{y}^{\prime}L/k_{x}=L/k_{x}a$ at a distance $L$ in the $x$
direction. We see that the correlation length $\delta\sim1/\Delta q$ so that
both arguments will lead to the same resolution. In order to put the argument
on a firmer footing we need to show that a wave acquires a correlation length
$\delta$ on passing through a slit of width $a.$ $\delta$ is the length
associated with a Fraunhofer diffraction pattern of a slit. The problem is
then to show that this length can arise in the case of Fresnel diffraction
(finite distance $L$). This is the content of the van-Cittert Zernike theorem
(\cite{BW}, \cite{champ}, \cite{cz}) mentioned above. In this way we can
describe the scattering process in terms of diffraction at the defining slits
and the correlation length of the incident beam. We will be able to describe
the scattering in real space instead of in the Fourier transform space defined
by $\overrightarrow{q}$.

For a small angle scattering instrument with wavelength of $10~$%
\emph{Angst.}\ and collimating slits $2cm$ wide separated by $L=10m$ the
correlation width will be about $1600~$\emph{Angst.,}\ \emph{i.e. }much larger
than the wavelength but much smaller than the slit width. Correlations over
larger distances than this cannot be observed with the assumed instrument.

Applying the same argument to a chopper for massive particles we see that for
a chopper opening $T$ and time of flight $t_{o}$ we will have $\Delta
E/E=2T/t_{o}=\Delta\omega/\omega,~\left(  \hbar\omega=E=mv^{2}/2\right)  $ by
the usual classical arguments. Quantum mechanically, according to the
uncertainty principle, passing through a chopper of width $T$ will result in a
spread of energy given by $\Delta\omega^{\prime}\thicksim1/T$ and this spread
in energy will result in a spread in arrival times $\Delta t/t_{o}%
\thicksim\Delta v/v\thicksim\Delta\omega^{\prime}/2\omega\thicksim1/2\omega
T$. Thus we should expect a correlation time of $\Delta t\thicksim
t_{o}/2\omega T\sim1/\Delta\omega$. For a time of flight instrument with pulse
width $T=10^{-5}s$ and a flight path of $L=5m$, neutron wavelength
$\lambda=5~$\emph{Angst.}\ we obtain $\Delta t\thicksim30ps$ and this is the
limit to the correlation times which can be observed with such an instrument.

\smallskip

\section{A new look at scattering from many particle systems}

\label{sec2}We will now show how the effects discussed above are contained in
a quantum-mechanical calculation of the scattering. Following the usual
treatment of scattering \cite{ns}, sec. 2.1, \cite{lips} the scattered wave is
given by%

\begin{equation}
\psi_{sc}\left(  \overrightarrow{r},t\right)  =\int G_{o}\left(
\overrightarrow{r}-\overrightarrow{r_{s}},t-t_{s}\right)  V\left(
\overrightarrow{r_{s}},t_{s}\right)  \psi_{sc}\left(  \overrightarrow{r_{s}%
},t_{s}\right)  d^{3}r_{s}dt_{s} \label{one}%
\end{equation}
where $\psi_{sc}\left(  \overrightarrow{r_{s}}\right)  $ is the exact solution
for the wave function at the position $\overrightarrow{r_{s}}$, and $G_{o}$ is
the Green's function for the unperturbed problem $\left(  V=0\right)  $. The
notation is explained in figs. 4 and 6. As is usual in van Hove scattering we
make the Born approximation by replacing $\psi_{sc}\left(  \overrightarrow
{r_{s}}\right)  $ by the incident wave at the position of the scattering
sample, $\psi_{in}\left(  \overrightarrow{r_{s}}\right)  $ and take (for the
case of neutron scattering)
\begin{equation}
V\left(  \overrightarrow{r_{s}},t\right)  =\sum_{i}\frac{2\pi\hbar^{2}}%
{m}b_{i}\delta\left(  \overrightarrow{r_{s}}-\overrightarrow{r_{i}}\left(
t\right)  \right)  \equiv\frac{2\pi\hbar^{2}}{m}\overline{b}\rho\left(
\overrightarrow{r_{s}},t\right)  \label{Vferm}%
\end{equation}

(for coherent scattering), \emph{i.e. }the Fermi pseudo potential where
$\overrightarrow{r_{i}}$ is the position of the $i^{th}$ nucleus, the sum is
over all nuclei in the scattering system and $\rho\left(  \overrightarrow
{r_{s}},t\right)  $ as defined in (\ref{Vferm}) is the particle density of the
scatterer.\footnote{In the interests of simplicity we omit prefactors and
constants in the body of this work concentrating on those terms which are
essential to the physics. We give a complete treatment in the appendix where
we also show how to extract the cross section from our results.}%

\begin{equation}
\psi_{sc}\left(  \overrightarrow{r},t\right)  =\int G_{o}\left(
\overrightarrow{r}-\overrightarrow{r_{s}},t-t_{s}\right)  V\left(
\overrightarrow{r_{s}},t_{s}\right)  \psi_{in}\left(  \overrightarrow{r_{s}%
},t_{s}\right)  d^{3}r_{s}dt_{s} \label{two}%
\end{equation}

The integration is carried out over the sample volume and the time interval
during which the sample is exposed to the beam.

Equation (\ref{two}) describes the scattered wave as a superposition of waves
produced by many scattering events each occurring at the different space time
points $\left(  \overrightarrow{r_{s}},t_{s}\right)  $(Fig. 1). This is in
contrast to the usual van Hove treatment \cite{vanh}, where one begins with
the Fermi golden rule for time dependent perturbations and shows, replacing
the energy conservation delta function by its Fourier transform, that the
probability of scattering is given by the space-time Fourier transform of the
correlation function of $V\left(  \overrightarrow{r_{s}},t_{s}\right)  $. In
the appendix to \cite{nsplit} we have shown how one can derive the van Hove
result in a more direct way. Introducing $\overrightarrow{r_{s}}%
-\overrightarrow{r_{s}}^{\prime}=\overrightarrow{\delta\text{ }}$,
$t_{s}-t_{s}^{\prime}=\tau$ we have from (\ref{two}) for the intensity at the
detector, $\left(  \overrightarrow{r_{d}},t_{d}\right)  $:%

\begin{align}
\left|  \psi_{sc}\left(  \overrightarrow{r_{d}},t_{d}\right)  \right|  ^{2}
&  =\int d^{3}r_{s}dt_{s}\int d^{3}r_{s}^{\prime}dt_{s}^{\prime}G_{o}\left(
\overrightarrow{r_{d}}-\overrightarrow{r_{s}},t_{d}-t_{s}\right)  G_{o}%
^{*}\left(  \overrightarrow{r_{d}}-\overrightarrow{r_{s}^{\prime}},t_{d}%
-t_{s}^{\prime}\right)  \times\nonumber\\
&  \times V\left(  \overrightarrow{r_{s}},t_{s}\right)  V^{*}\left(
\overrightarrow{r_{s}^{\prime}},t_{s}^{\prime}\right)  \psi_{in}\left(
\overrightarrow{r_{s}},t_{s}\right)  \psi_{in}^{*}\left(  \overrightarrow
{r_{s}^{\prime}},t_{s}^{\prime}\right) \nonumber\\
&  =\int d^{3}r_{s}dt_{s}\int d^{3}\delta d\tau G_{o}\left(  \overrightarrow
{r_{d}}-\overrightarrow{r_{s}},t_{d}-t_{s}\right)  G_{o}^{*}\left(
\overrightarrow{r_{d}}-\overrightarrow{r_{s}}+\overrightarrow{\delta\text{ }%
},t_{d}-t_{s}+\tau\right) \nonumber\\
&  \times V\left(  \overrightarrow{r_{s}},t_{s}\right)  V^{*}\left(
\overrightarrow{r_{s}}-\overrightarrow{\delta\text{ }},t_{s}-\tau\right)
\psi_{in}\left(  \overrightarrow{r_{s}},t_{s}\right)  \psi_{in}^{*}\left(
\overrightarrow{r_{s}}-\overrightarrow{\delta\text{ }},t_{s}-\tau\right)
\end{align}

Substituting the Fermi potential we find%

\begin{align}
\left|  \psi_{sc}\left(  \overrightarrow{r_{d}},t_{d}\right)  \right|  ^{2}  &
=\int d^{3}r_{s}dt_{s}\int d^{3}\delta d\tau G_{o}\left(  \overrightarrow
{r_{d}}-\overrightarrow{r_{s}},t_{d}-t_{s}\right)  G_{o}^{\ast}\left(
\overrightarrow{r_{d}}-\overrightarrow{r_{s}}+\overrightarrow{\delta\text{ }%
},t_{d}-t_{s}+\tau\right)  \label{24}\\
& \times G_{s}\left(  \overrightarrow{\delta\text{ }},\tau\right)
R_{in}\left(  \overrightarrow{\delta\text{ }},\tau\right)
\end{align}

Here
\begin{equation}
G_{s}\left(  \overrightarrow{\delta\text{ }},\tau\right)  =\left\langle
\rho\left(  \overrightarrow{r_{s}},t_{s}\right)  \rho\left(  \overrightarrow
{r_{s}}-\overrightarrow{\delta\text{ }},t_{s}-\tau\right)  \right\rangle _{s}
\label{Gs}%
\end{equation}
is the (van Hove) density-density correlation function and is independent of
$\overrightarrow{r_{s}}$, $t_{s}$ for the usual case of a homogeneous system.
The result depends on the pair correlation function (\ref{Gs}) due to the fact
that the intensity is proportional to the square of the amplitude as can
readily be seen in the above derivation. Because we restrict ourselves to the
Born approximation the result is independent of higher order correlation functions.%

\begin{equation}
R_{in}\left(  \overrightarrow{\delta\text{ }},\tau\right)  =\left\langle
\psi_{in}\left(  \overrightarrow{r_{s}},t_{s}\right)  \psi_{in}^{*}\left(
\overrightarrow{r_{s}}-\overrightarrow{\delta\text{ }},t_{s}-\tau\right)
\right\rangle _{in}%
\end{equation}
is the auto-correlation function of the incident beam (See fig.2 for the time
independent case.). The brackets $\left\langle {}\right\rangle _{s}$ and
$\left\langle {}\right\rangle _{in}$ indicate statistical averages over the
sample and incoming beam ensembles respectively.

$R_{in}\left(  \overrightarrow{\delta\text{ }},\tau\right)  $ will be seen to
be responsible for the contributions of the incident beam to the resolution of
the measurement, as, according to (\ref{24}), the intensity only depends on
the correlation function $G_{s}$ at values of $\left(  \overrightarrow
{\delta\text{ }},\tau\right)  $ where $R_{in}$ is significant.

The wave function incident on the sample $\psi_{in}$, is given by%

\begin{equation}
\psi_{in}\left(  \overrightarrow{r_{s}},t_{s}\right)  =\int d^{2}r_{o}%
dt_{o}G_{o}\left(  \overrightarrow{r_{s}}-\overrightarrow{r_{o}},t_{s}%
-t_{o}\right)  \psi_{o}\left(  \overrightarrow{r_{o}},t_{o}\right)
\end{equation}
\medskip where $\psi_{o}\left(  \overrightarrow{r_{o}},t_{o}\right)  $ is the
incident wave function at the entrance of the apparatus\footnote{Strictly
speaking we have $\psi_{in}\left(  \overrightarrow{r_{s}},t_{s}\right)  =\int
d^{2}r_{o}dt_{o}\frac{\partial\widetilde{G}_{o}\left(  \overrightarrow{r_{s}%
}-\overrightarrow{r_{o}},t_{s}-t_{o}\right)  }{\partial\overrightarrow{n}}%
\psi_{o}\left(  \overrightarrow{r_{o}},t_{o}\right)  $ where $\widetilde
{G}_{o}$is a Green's function satisfying the boundary condition $\widetilde
{G}_{o}=0$ on the surface of integration, and $\partial/\partial
\overrightarrow{n}$ is the gradient in the direction of the surface normal
\cite{mf},\cite{ggtdo}. However this only introduces a change in the
prefactor. See the appendix for a complete treatment.}. We then find%

\begin{equation}
R_{in}\left(  \overrightarrow{\delta\text{ }},\tau\right)  =\int d^{2}%
r_{o}dt_{o}G_{o}\left(  \overrightarrow{r_{1}},t_{s}-t_{o}\right)  G_{o}%
^{*}\left(  \overrightarrow{r_{1}}-\overrightarrow{\delta\text{ }},t_{1}%
-\tau\right)  \label{rin}%
\end{equation}
where $\overrightarrow{r_{1}}=\overrightarrow{r_{s}}-\overrightarrow{r_{o}}$ ;
$t_{1}=t_{s}-t_{o}$ and we have made the assumption that the incident beam is
completely uncorrelated, \emph{i.e. }$\left\langle \psi_{o}\left(
\overrightarrow{r_{o}},t_{o}\right)  \psi_{o}^{*}\left(  \overrightarrow
{r_{o}}^{\prime},t_{o}^{\prime}\right)  \right\rangle _{in}=\delta^{3}\left(
\overrightarrow{r_{o}}-\overrightarrow{r_{o}}^{\prime}\right)  \delta\left(
t_{o}-t_{o}^{\prime}\right)  $ as is usual in the derivation of the van
Cittert-Zernike theorem \cite{BW},\cite{champ} and \cite{cz}. By the
Wiener-Khintchine theorem \cite{champ} we see this is equivalent to $\left|
A\left(  \overrightarrow{k}\right)  \right|  ^{2}=const$, \emph{i.e. }the
incident spectrum is broad compared to that selected by the instrument.

In a scattering experiment we measure the integral of $\left|  \psi
_{sc}\left(  \overrightarrow{r_{d}},t_{d}\right)  \right|  ^{2}$ over the
entrance area of the detector and, in the case of a time of flight experiment,
a definite time interval, $dt_{d}$. From (\ref{24}) the intensity at the
detector, $I_{d}$, is given by%

\begin{equation}
\int d^{2}r_{d}dt_{d}\left|  \psi_{sc}\left(  \overrightarrow{r_{d}}%
,t_{d}\right)  \right|  ^{2}=\int d^{3}r_{s}dt_{s}\int d^{3}\delta d\tau
R_{out}\left(  \overrightarrow{\delta\text{ }},\tau,\overrightarrow{r_{2}%
},t_{2}\right)  G_{s}\left(  \overrightarrow{\delta\text{ }},\tau\right)
R_{in}\left(  \overrightarrow{\delta\text{ }},\tau,\overrightarrow{r_{1}%
},t_{1}\right)  \label{biggy}%
\end{equation}
where%

\begin{equation}
R_{out}\left(  \overrightarrow{\delta\text{ }},\tau,\overrightarrow{r_{2}%
},t_{2}\right)  =\int d^{2}r_{d}dt_{d}G_{o}\left(  \overrightarrow{r_{2}%
},t_{2}\right)  G_{o}^{*}\left(  \overrightarrow{r_{2}}+\overrightarrow
{\delta\text{ }},t_{2}+\tau\right)  \label{rout}%
\end{equation}
and $\overrightarrow{r_{2}}=\overrightarrow{r_{d}}-\overrightarrow{r_{s}}$,
$t_{2}=t_{d}-t_{s}$.

Although (\ref{biggy}) contains an explicit dependence on only space and time
variables, the dependence on the energy and momentum transfer will be seen to
emerge from the variation of the phases of the $G_{o}$ functions.

Now we see there is a symmetry between the integral over the detector
parameters, $\left(  \overrightarrow{r_{d}},t_{d}\right)  $, of the product of
Green's functions depending on $\left(  \overrightarrow{r_{2}},t_{2}\right)
$, $R_{out}\left(  \overrightarrow{\delta\text{ }},\tau\right)  $, (equ.
\ref{rout}) and the integral over the input slit $\left(  \overrightarrow
{r_{o}},t_{o}\right)  $ of the Green's function product depending on $\left(
\overrightarrow{r_{1}},t_{1}\right)  $, $R_{in}\left(  \overrightarrow
{\delta\text{ }},\tau\right)  $, (equ. \ref{rin}) so that we only have to do
the calculation of one of the pairs, say the latter. We then obtain the former
integral by applying the symmetry rules:%

\begin{equation}
\overrightarrow{\delta\text{ }}\rightarrow-\overrightarrow{\delta\text{ }%
},\;\tau\rightarrow-\tau,\;t_{2}\rightarrow t_{1},\;\overrightarrow{r_{2}%
}\rightarrow\overrightarrow{r_{1}},\;\overrightarrow{r_{s}}\rightarrow
-\overrightarrow{r_{s}},\;t_{s}\rightarrow-t_{s}. \label{symrule}%
\end{equation}

While functions such as $R_{in,out}\left(  \overrightarrow{\delta\text{ }%
},\tau\right)  $ are usually called 'correlation functions' they really
represent a loss of correlation as their arguments increase. In the present
context it is instructive to keep this in mind by referring to them as '(loss
of) correlation functions'.

Thus while $R_{in}\left(  \overrightarrow{\delta\text{ }},\tau,\overrightarrow
{r_{1}},t_{1}\right)  $ represents the loss of correlation between pairs of
rays starting at the entrance slit $\left(  \overrightarrow{r}_{o}\right)  $,
propagating to neighboring space time points in the sample $\left(
\overrightarrow{r_{s}},t_{s}\right)  $, $\left(  \overrightarrow{r_{s}%
}^{\prime},t_{s}^{\prime}\right)  $, $R_{out}\left(  \overrightarrow
{\delta\text{ }},\tau\right)  $ represents the loss of correlation due to the
difference in path lengths in propagating from two neighboring points in the
sample $\left(  \overrightarrow{r_{s}},t_{s}\right)  $, $\left(
\overrightarrow{r_{s}}^{\prime},t_{s}^{\prime}\right)  $ to the detector
$\left(  \overrightarrow{r}_{d},t_{d}\right)  $. $G_{s}\left(  \overrightarrow
{\delta\text{ }},\tau\right)  $ is the loss of correlation between the points
$\left(  \overrightarrow{r_{s}},t_{s}\right)  $, $\left(  \overrightarrow
{r_{s}}^{\prime},t_{s}^{\prime}\right)  $ due to the internal dynamics of the
scattering system (See fig. 3). Implicit in the above discussion is the fact
that the detector responds to the value of $\left|  \psi_{sc}\right|  ^{2}$ at
a single point integrated over the detector area.

The above discussion brings out the fact that the scattering process is
intrinsically an interference phenomenon, the cross section being a measure of
the loss of correlation between waves following paths separated by $\left(
\overrightarrow{\delta\text{ }},\tau\right)  $ at the sample.

In this paper we limit ourselves to pure coherent scattering. The separation
of the scattering into a coherent and an incoherent part can be carried out
exactly as in the conventional treatment. \cite{vanh}, \cite{ns}. Also in the
interests of simplicity we choose to work with a real 'hydrodynamic' density
function $\rho\left(  \overrightarrow{r},t\right)  $. Replacement of this by
$\sum_{i}\delta\left(  \overrightarrow{r}-\overrightarrow{r}_{i}(t)\right)  $
yields results consistent with the conventional approach \cite{vanh}, (see
also \cite{nsplit}).

\section{Elastic Scattering in Real Space}

\label{elscatt}In this section we specialize to the case of scattering from a
static system. We will consider time dependent systems in the following sections.

Using the known form of the Green's function for the time independent
Schr\"{o}dinger equation,
\[
G_{o}=e^{ik\left|  \overrightarrow{r}-\overrightarrow{r_{s}}\right|  }/\left|
\overrightarrow{r}-\overrightarrow{r_{s}}\right|
\]
we calculate the input product of Green's functions%

\begin{equation}
\sim e^{ik\left(  \left|  \overrightarrow{r}_{s}-\overrightarrow{r_{o}%
}\right|  -\left|  \overrightarrow{r}_{s}^{\prime}-\overrightarrow{r_{o}%
}\right|  \right)  }%
\end{equation}
by expanding%

\begin{equation}
\left|  \overrightarrow{r}_{s}-\overrightarrow{r}_{o}\right|  -\left|
\overrightarrow{r}_{s}^{\prime}-\overrightarrow{r}_{o}\right|  =\frac
{\overrightarrow{\delta}\cdot\left(  \overrightarrow{r}_{s}-\overrightarrow
{r}_{o}\right)  }{\left|  \overrightarrow{r}_{s}-\overrightarrow{r}%
_{o}\right|  }\approx\frac{\overrightarrow{\delta}\cdot\left(  \overrightarrow
{r}_{s}-\overrightarrow{r}_{o}\right)  }{\left|  \overrightarrow{r}%
_{s}\right|  }\approx\frac{\overrightarrow{\delta}\cdot\left(  \overrightarrow
{r}_{s}-\overrightarrow{r}_{o}\right)  }{\left|  \overrightarrow{\underline
{r}}_{1}\right|  }%
\end{equation}
which holds to first order in $\overrightarrow{\delta}=\overrightarrow{r}%
_{s}-\overrightarrow{r}_{s}^{\prime}$. $\overrightarrow{\underline{r}}%
_{1}=\overrightarrow{\underline{r}}_{s}$ is the location of the center of the
sample and we rewrite $\overrightarrow{r}_{s}\Rightarrow\overrightarrow
{\underline{r}}_{s}+\overrightarrow{\epsilon}_{s}$ where $\overrightarrow
{\epsilon}_{s}$ is now a small quantity (see fig. 4). Then%

\begin{align}
R_{in}\left(  \overrightarrow{\delta}\right)   &  =e^{ik\frac{\overrightarrow
{\delta}\cdot\left(  \overrightarrow{\underline{r}}_{s}+\overrightarrow
{\epsilon}_{s}\right)  }{r_{1}}}\int d^{2}r_{o}e^{-ik\frac{\overrightarrow
{\delta}\cdot\overrightarrow{r}_{o}}{r_{1}}}=e^{ik\frac{\overrightarrow
{\delta}\cdot\left(  \overrightarrow{\underline{r}}_{s}+\overrightarrow
{\epsilon}_{s}\right)  }{r_{1}}}\int_{-a}^{a}dy_{o}e^{-ik\frac{\delta_{y}%
y_{o}}{r_{1}}}=\nonumber\\
&  =e^{i\overrightarrow{\underline{k}}_{1}\cdot\overrightarrow{\delta}%
}e^{ik\frac{\overrightarrow{\delta}\cdot\overrightarrow{\epsilon}_{s}}{r_{1}}%
}2a\left(  \frac{\sin k\frac{\delta_{y}a}{r_{1}}}{k\frac{\delta_{y}a}{r_{1}}%
}\right)  \label{rinsas}%
\end{align}
where we have taken a one dimensional slit of width $2a$ and $\overrightarrow
{\underline{k}}_{1}=k$ $\overrightarrow{\underline{r}}_{1}/r_{1}$ and
$r_{i}=\left|  \overrightarrow{\underline{r}}_{i}\right|  $. $\delta_{y}$ is
the component of $\overrightarrow{\delta}$ perpendicular to the incident beam
(see fig.4).

The result (\ref{rinsas}) is known as the van Cittert-Zernike theorem,
\cite{BW}, \cite{champ}, \cite{cz} and has the form of the Fraunhofer formula.
Note that this formula is applicable to geometries where one would not
normally observe a Fraunhofer diffraction pattern, except in the case of a
spherical wave converging on the point of observation. In applications to
optics the phase factors in (\ref{rinsas}), depending on the position of the
pair of points with respect to the center of the sample, as well as on the
distance between them are often neglected so that the correlation functions
can be considered as real functions depending only on the distance between the
two points. However the phase factors play a crucial role in scattering,
bringing in the influence of the sample size on the resolution and the
dependence on the energy and momentum transfer, as shown below.

By symmetry (\ref{symrule}) we can write%

\begin{equation}
R_{out}\left(  \overrightarrow{\delta\text{ }}\right)  =e^{-i\overrightarrow
{\underline{k}}_{2}\cdot\overrightarrow{\delta}}e^{ik\frac{\overrightarrow
{\delta}\cdot\overrightarrow{\epsilon}_{s}}{r_{2}}}2d\left(  \frac{\sin
k\frac{\delta_{y}^{\prime}d}{r_{2}}}{k\frac{\delta_{y}^{\prime}d}{r_{2}}%
}\right)  \label{routsas}%
\end{equation}
where $\delta_{y}^{\prime}$ represents the component of $\overrightarrow
{\delta\text{ }}$ in the direction perpendicular to the direction of the
scattered beam, $\overrightarrow{\underline{k}}_{2}=k\overrightarrow
{\underline{r}}_{2}/r_{2}\,$and $d$ is the width of the detector slit (see
fig. 4).

\smallskip Then writing the steady state form of (\ref{biggy}) we have%

\begin{equation}
I_{d}=\int d^{3}\epsilon_{s}\int d^{3}\delta R_{out}\left(  \overrightarrow
{\delta\text{ }},\overrightarrow{r_{2}}\right)  G_{s}\left(  \overrightarrow
{\delta\text{ }}\right)  R_{in}\left(  \overrightarrow{\delta\text{ }%
},\overrightarrow{r_{1}}\right)  \label{biggyss}%
\end{equation}

Substituting (\ref{rinsas}) and (\ref{routsas}) in (\ref{biggyss}) we have%

\begin{equation}
I_{d}=\int d^{3}\epsilon_{s}\int d^{3}\delta e^{-i\overrightarrow
{\underline{q}}\cdot\overrightarrow{\delta}}e^{ik\frac{\overrightarrow{\delta
}\cdot\overrightarrow{\epsilon}_{s}}{r^{\prime}}}2d\left(  \frac{\sin
k\frac{\delta_{y^{\prime}}d}{r_{2}}}{k\frac{\delta_{y^{\prime}}d}{r_{2}}%
}\right)  G_{s}\left(  \overrightarrow{\delta\text{ }}\right)  2a\left(
\frac{\sin k\frac{\delta_{y}a}{r_{1}}}{k\frac{\delta_{y}a}{r_{1}}}\right)
;\qquad\frac{1}{r^{\prime}}=\frac{1}{r_{1}}+\frac{1}{r_{2}} \label{36}%
\end{equation}
The term $\exp\left(  ik\frac{\overrightarrow{\delta}\cdot\overrightarrow
{\epsilon}_{s}}{r^{\prime}}\right)  $ comes from the phase factor which is
normally neglected in optics. (\emph{Note that }$\overrightarrow{\underline
{q}}=\overrightarrow{\underline{k}}_{2}-\overrightarrow{\underline{k}}_{1}$, a
point which will be discussed in the next section.) In the present case this
factor means that moving the location of the correlation regions defined by
the two $\sin x/x$ functions (varying $\overrightarrow{\epsilon}_{s}$) in
(\ref{36}) causes an additional difference in the two optical paths going from
the entrance slit to the detector slit via the two points separated by
$\overrightarrow{\delta}$, \emph{i.e.} the two paths in fig. 4. This optical
path difference represents an additional loss in correlation as the pair of
points separated by a constant $\overrightarrow{\delta}$ move away from the
optical axis and determines the influence of the sample slit width $\left(
2b\right)  $ on the resolution.

\subsection{Small angle scattering}

For simplicity we will now consider the case of small angle scattering where
the only significant component of $\overrightarrow{\delta}$ is $\delta
_{y}=y_{s}-y_{s}^{\prime}$ and $\delta_{y}\approx\delta_{y^{\prime}}.$

Performing the $\epsilon_{s}$ integration in (\ref{36}) over a slit of width
$2b$ $\left(  \int_{-b}^{b}d\epsilon_{s}\right)  $ defining the sample size yields%

\begin{equation}
I_{d}=\int d^{3}\delta e^{-i\overrightarrow{\underline{q}}\cdot\overrightarrow
{\delta}}G_{s}\left(  \overrightarrow{\delta\text{ }}\right)  2d\left[
\frac{\sin k\frac{\delta_{y^{\prime}}d}{r_{2}}}{k\frac{\delta_{y^{\prime}}%
d}{r_{2}}}\right]  2b\left[  \frac{\sin k\frac{\delta_{y}b}{r^{\prime}}%
}{k\frac{\delta_{y}b}{r^{\prime}}}\right]  2a\left[  \frac{\sin k\frac
{\delta_{y}a}{r_{1}}}{k\frac{\delta_{y}a}{r_{1}}}\right]  \label{xx}%
\end{equation}
where we have neglected the extension of the sample in the $z$ direction.

For $a,b,d$ small, corresponding to a high $\overrightarrow{q}$ resolution,
the functions in square brackets approach unity for the values of $\delta$ for
which $G_{s}$ is significant, and we have the usual result%

\begin{equation}
I_{d}\sim S\left(  \overrightarrow{q}\right)  =\int d^{3}\delta
e^{-i\overrightarrow{q}\cdot\overrightarrow{\delta}}G_{s}\left(
\overrightarrow{\delta}\right)  \label{norm}%
\end{equation}

The same result is obtained if we take the incident and outgoing states as
plane waves. The result of an experiment for non-negligible $a,b,d$ must then
be found by taking the convolution of (\ref{norm}) with an instrumental
resolution function. In the approach of (\ref{xx}) however, the resolution is
contained in the (loss of) correlation functions in the square brackets. Equ.
(\ref{xx}) clearly shows how correlations between points in the sample
separated by larger values of $\delta$ contribute less to the intensity.
However due to the nature of the $\sin x/x$ functions there are some
contributions to the scattering from these larger values, and some regions of
$\delta$ contribute with a negative sign. These effects, at the 10\% level,
which do not appear if we approximate the $\sin x/x$ functions by Gaussians of
an appropriate width as is often done in practice, become more important for
measurements at poor resolution such as in quasi-elastic line broadening.

\subsubsection{Comparison with conventional treatment}

We see that equ. (\ref{xx}) is the Fourier transform with respect to
$\overrightarrow{\delta}$ of a product of $G\left(  \overrightarrow{\delta
}\right)  $ with a function $H\left(  a,b,\delta_{y}\right)  $ which is itself
the product of three functions,$h_{2}\left(  d,\delta_{y}^{\prime}\right)  $,
$h^{\prime}\left(  b,\delta_{y}\right)  $and $h_{1}\left(  a,\delta
_{y}\right)  $. Thus the result (\ref{xx}) is the convolution of the Fourier
transform of $G\left(  \overrightarrow{\delta}\right)  $, \emph{i.e. }
$S\left(  \overrightarrow{q}\right)  $, with the Fourier transform of $H$. The
latter is the convolution of the Fourier transforms of each of the $h$ functions%

\begin{equation}
\eta_{a}\left(  q_{y}\right)  =\int d^{3}\delta e^{-i\overrightarrow{q}%
\cdot\overrightarrow{\delta}}h_{1}\left(  a,\delta_{y}\right)  =\frac{r_{1}%
}{k}\left[  u\left(  q_{y}+\frac{ka}{r_{1}}\right)  -u\left(  q_{y}-\frac
{ka}{r_{1}}\right)  \right]  \label{eta}%
\end{equation}
($r_{1}/k$ is a normalizing factor so that $\eta_{a}\left(  q_{y}\right)  $ is
normalized to $2a$ and $u\left(  x\right)  $ is the unit step function). In
$\eta_{b}\left(  q_{y}\right)  $ the term $\left(  a/r_{1}\right)  $ is
replaced by $\left(  b/r^{\prime}\right)  $ and in $\eta_{d}\left(
q_{y}\right)  $ it is replaced by $\left(  d/r_{2}\right)  $. The convolution
of two rectangular functions in $q_{y},$ $\eta_{a}$ and $\eta_{b}$, yields a
trapezoid with a base bounded by
\[
\left|  q_{y}\right|  <k\left(  a/r_{1}+b\left(  1/r_{1}+1/r_{2}\right)
\right)
\]
and a top bounded by
\[
\left|  q_{y}\right|  <k\left|  a/r_{1}-b\left(  1/r_{1}+1/r_{2}\right)
\right|  ,
\]
just what is expected on geometrical grounds for a collimator consisting of
two slits of width $2a$ and $2b$ separated by a distance $x$ as seen by a
point $r_{d}$ away from the second slit (see fig. 5). The subsequent
convolution with $\eta_{d}\left(  q_{y}\right)  $ yields the effect of finite
detector width. From (\ref{xx}) the correlation length of the input function
$h_{1}$ is seen to be $y_{c}=r_{1}/ka$, while (\ref{eta}) gives the width
$y_{q}$ in $q$, to be $y_{q}=ka/r_{1}$. Hence $y_{c}\cdot y_{q}=1$ $or$
$y_{c}\cdot y_{p}=\hbar,$ where the momentum width $y_{p}=\hbar y_{q}$.

Thus our result (\ref{xx}) is identical with the usual treatment which yields
the convolution of $S\left(  \overrightarrow{q}\right)  $ with the resolution
function but our view of the scattering is quite different. We see the
scattering process as the interaction with the sample of a state whose wave
function has a certain spatial auto-correlation function determined by the
optical properties of the devices used to define the incoming beam. This has
the effect that only pairs of scattering points whose separation is less than
the correlation length of the incident beam contribute significantly to the
scattered wave and pairs with different separations $\overrightarrow{\delta}$
are weighted by the auto-correlation function of the incoming beam evaluated
at $\overrightarrow{\delta}$. Analogous statements apply to the outgoing beam.

Similar considerations have been used by many people (\emph{e.g.
}\cite{pynnco}) as a means of deconvoluting the instrument resolution from the
measured data. However, in the work cited no mention is made of the physical
significance of the Fourier transform of the instrument resolution function
(our $H\left(  \delta\right)  $ is called $R\left(  k\right)  $ in
(\cite{pynnco})) as the (loss of) correlation functions of the incident and
scattered wave functions. In the above treatment we see how the resolution
function is the result of the optical properties of the devices defining the
incident and scattered beams.

\subsubsection{{}Discussion of the result for small angle scattering}

In the previous section we have seen that the contribution to $\left\langle
\left|  \psi_{sc}\left(  \overrightarrow{r_{d}}\right)  \right|
^{2}\right\rangle $ from the correlation function of the incident beam
represents the effect on the resolution due to the collimation of the incident
beam and the correlation function of the scattered beam results in that part
of the resolution due to the angular resolution of the scattered beam. Each
element of the optical system defining the beam is seen to contribute a factor
to the overall correlation function.

In the following we try to illustrate the relation between our treatment of
the scattering problem and the conventional approach by concentrating on the
contribution of the incident beam to the resolution. If we write%

\begin{equation}
\Psi\,_{inc}\left(  \overrightarrow{r_{s}}\right)  =\int d^{3}k_{1}A\left(
\overrightarrow{k}_{1}\right)  e^{i\overrightarrow{k_{1}}\cdot\overrightarrow
{r_{s}}}%
\end{equation}

then we have
\begin{align}
\left\langle \Psi\,_{inc}\left(  \overrightarrow{r_{s}}\right)  \Psi_{inc}%
^{*}\left(  \overrightarrow{r_{s}^{\prime}}\right)  \right\rangle _{b}  &
=\int d^{3}r_{s}\int d^{3}k_{1}\int d^{3}k_{1}^{^{\prime}}\left\langle
A\left(  k_{1}\right)  A^{*}\left(  k_{1}^{^{\prime}}\right)  \right\rangle
e^{i\overrightarrow{k_{1}}\cdot\overrightarrow{r_{s}}-\overrightarrow
{k_{1}^{^{\prime}}}\cdot\overrightarrow{r_{s}^{^{\prime}}}}\nonumber\\
&  =\int d^{3}k_{1}\left\langle \left|  A\left(  k_{1}\right)  \right|
^{2}\right\rangle e^{i\overrightarrow{k_{1}}\cdot\left(  \overrightarrow
{r_{s}}-\overrightarrow{r_{s}^{^{\prime}}}\right)  }%
\end{align}

Substituting this into (\ref{biggyss}) and writing $\psi_{out}%
=e^{i\overrightarrow{k}_{2}\cdot\overrightarrow{r}}$ we obtain%

\begin{equation}
\int d^{3}\epsilon_{s}\int d^{3}k_{1}\left\langle \left|  A\left(
k_{1}\right)  \right|  ^{2}\right\rangle \int d^{3}\delta e^{-i\overrightarrow
{q}\cdot\overrightarrow{\delta}}G_{s}\left(  \overrightarrow{\delta}\right)
\sim\int d^{3}k_{1}\left\langle \left|  A\left(  k_{1}\right)  \right|
^{2}\right\rangle S\left(  \overrightarrow{q}\right)  \label{squatty}%
\end{equation}

In this case we have taken $\overrightarrow{q}=\overrightarrow{k_{2}%
}-\overrightarrow{k_{1}}$ and we obtain $S\left(  \overrightarrow{q}\right)  $
averaged over the momentum spectrum of the incident beam.

In the previous section we took $\overrightarrow{\underline{q}}%
=\overrightarrow{\underline{k}}_{2}-\overrightarrow{\underline{k}_{1}}$ ,
\emph{i.e.} the difference between the \underline{nominal} final wave vector,
and the \underline{nominal} incoming wave vector. By this we mean the wave
vector corresponding to a central ray of the incident or scattered beam. Then,
in that case, the distribution in $\overrightarrow{k_{1}},$ or what contains
the same information, the correlation function of the incident beam produced
by the optical properties of the defining slits, gives the usual result of
$S\left(  \overrightarrow{q}\right)  $ convoluted with an instrumental
resolution function. On the other hand, in this section by taking%

\begin{equation}
\overrightarrow{q}=\overrightarrow{k_{2}}-\overrightarrow{k_{1}} \label{squat}%
\end{equation}
we have assumed perfect knowledge of $\overrightarrow{k_{1}},$
$\overrightarrow{k_{2}}$ and we complete the calculation (equ. \ref{squatty})
by integrating over all values of $\overrightarrow{q}$ corresponding to the
spread in $\overrightarrow{k_{1}}$ values in the incident beam. This will of
course yield the same results as the previous treatment.

\subsection{Elastic scattering at arbitrary angle}

We will assume a cylindrical shaped sample (radius $R_{s}$) with its axis
perpendicular to the scattering plane and thus consider the problem as a two
dimensional one. Then the integral over the sample volume in (\ref{36}) can be
written as%

\begin{align}
\int d^{3}\epsilon_{s}e^{ik\frac{\overrightarrow{\delta}\cdot\overrightarrow
{\epsilon}_{s}}{r^{\prime}}}  &  =l\int_{0}^{R_{s}}\epsilon_{s}d\epsilon
_{s}\int_{0}^{2\pi}d\theta e^{ik\frac{\delta\epsilon_{s}\cos\theta}{r^{\prime
}}}=l\int_{0}^{R_{s}}\epsilon_{s}d\epsilon_{s}2\pi J_{0}\left(  k\frac
{\delta\epsilon_{s}}{r^{\prime}}\right)  =\nonumber\\
&  =2\pi R_{s}^{2}\frac{J_{1}\left(  k\frac{\delta R_{s}}{r^{\prime}}\right)
}{k\frac{\delta R_{s}}{r^{\prime}}}l
\end{align}
where $l$ represents the dimension of the sample perpendicular to the
scattering plane. The final result is then%

\begin{equation}
\int d^{2}r_{d}\left|  \psi_{sc}\left(  \overrightarrow{r_{d}}\right)
\right|  ^{2}=\int d^{3}\delta e^{-i\overrightarrow{q}\cdot\overrightarrow
{\delta}}G_{s}\left(  \overrightarrow{\delta}\right)  \mathbb{H}\left(
\overrightarrow{\delta}\right)  \label{qquutt}%
\end{equation}
where%

\begin{equation}
\mathbb{H}\left(  \overrightarrow{\delta}\right)  =2d\left(  \frac{\sin
k\frac{\delta_{y}^{\prime}d}{r_{2}}}{k\frac{\delta_{y}^{\prime}d}{r_{2}}%
}\right)  2\pi R_{s}^{2}\left(  \frac{J_{1}\left(  k\frac{\delta R_{s}%
}{r^{\prime}}\right)  }{k\frac{\delta R_{s}}{r^{\prime}}}\right)  2a\left(
\frac{\sin k\frac{\delta_{y}a}{r_{1}}}{k\frac{\delta_{y}a}{r_{1}}}\right)
l^{3} \label{H}%
\end{equation}
is the Fourier transform of the resolution function. The terms depending on
$a$, $d$ are identical to those discussed above in connection with small angle
scattering (\ref{xx}). The Fourier transform of the term containing $R_{s}$ is
given by%

\begin{align}
\int\delta d\delta d\theta e^{iq\delta\cos\theta}\left(  \frac{J_{1}\left(
\alpha\delta\right)  }{\alpha\delta}\right)    & =2\pi\int d\delta
J_{0}\left(  q\delta\right)  \left(  \frac{J_{1}\left(  \alpha\delta\right)
}{\alpha}\right)  =\\
& =\left\{
\begin{array}
[c]{l}%
\frac{2\pi}{\alpha^{2}}=\left(  \frac{r^{\prime}}{kR_{s}}\right)
^{2}=const\quad\left\{  q<\frac{kR_{s}}{r^{\prime}}\right\}  \\
0\quad\left\{  q>\frac{kR_{s}}{r^{\prime}}\right\}
\end{array}
\right.
\end{align}

\subsubsection{Discussion}

Just as in the small angle case the result in the general case (\ref{qquutt})
is seen as the Fourier transform of the product of $G_{s}$ with a product of
functions, each representing the contribution to the resolution of an
individual element of the beam handling system.

Corresponding to the $\sin x/x$ functions in one dimension the function
$J_{1}\left(  x\right)  /x$ in two dimensions represents a square shaped
resolution in $q$ but in the latter case the resolution is a function of
$\left|  \overrightarrow{q}\right|  $ rather than a component of
$\overrightarrow{q}$ in the one dimensional cases.

The term involving $\left(  \delta_{y}a\right)  $ gives the influence of the
width of the entrance slit $\left(  2a\right)  $ on the resolution; it depends
on $r_{1}$ but not on $r_{2}$. The term in $\left(  \delta_{y^{\prime}%
}d\right)  $ gives the influence of the detector slit width $\left(
2d\right)  $ on the resolution and depends on $r_{2}$. The remaining term
gives the influence of the sample size and depends on $r^{\prime}$ because an
uncertainty in position at the sample effects the angular resolution in both arms.

The argument of each term can be written as $\left(  \delta/\delta
_{ci}\right)  $ where the $\delta_{ci}$ are correlation lengths corresponding
to the size of the correlations that can be observed with a given
installation. When any of the terms is small, so is the contribution to the
total integral, so that the only significant contribution comes from values of
$\overrightarrow{\delta}$ for which each term in (\ref{H}) is significant.
Thus the resolution is determined by the overlap of the different correlation
volumes and the measurement can be optimized by choosing all regions as
approximately equal.

Equ. (\ref{qquutt}) has been derived for the case of a single incident energy.
It is easy to see that if we took an incident spectrum constant in a region
$\pm\Delta K$ that the result will be an additional term $\sin x/x$ with
$x=\Delta K\left(  \delta_{1}-\delta_{2}\right)  $, where $\delta_{1,2} $ are
the components of $\overrightarrow{\delta}$ along the directions of $k_{1,2}$
respectively. A more detailed discussion including the influence of the
scattering angle will be presented in a forthcoming paper.

\subsection{Neutron spin echo for elastic scattering.}

In neutron spin echo for elastic scattering, the relative phase shift between
the two spin states is given by $\overrightarrow{q}\cdot\overrightarrow
{\delta_{\text{NSE}}}$ with $\overrightarrow{q}=\overrightarrow{k}%
_{2}-\overrightarrow{k}_{1}$ \cite{nsplit}. Then the average beam polarization
will be given by equ. (\ref{squatty}) with $\overrightarrow{\delta}$ replaced
by $\left(  \overrightarrow{\delta}-\overrightarrow{\delta_{\text{NSE}}%
}\right)  $ and to this order the result is independent of the momentum
distribution of the incident beam. The dependence of the spin echo signal on
the incident beam comes from the fact that $\overrightarrow{\delta
_{\text{NSE}}}$ is a function of $\overrightarrow{k_{1}}$. We now show this in
more detail.

In \cite{nsplit} we have described various forms of spin echo spectrometer
from a space- time point of view. According to this picture the first arm of
the spectrometer causes the beam incident on the scattering sample to be split
in space and time, and this splitting results in the scattered beam sampling
the scattering system at points separated by this splitting. In this section
we show how the current view point yields the same results.

We note that we are considering only scattering with $\overrightarrow
{q}=\widehat{i}q_{y}$. In the case of spin echo the wave function:%

\begin{equation}
\Psi\,_{inc}\left(  \overrightarrow{r}\right)  =\int d^{3}k_{1}A\left(
k_{1}\right)  e^{i\overrightarrow{k_{1}}\cdot\overrightarrow{r}}%
\end{equation}
represents the beam arriving at the sample in the absence of the NSE magnetic
field. Turning on the first spin echo field (length $L$) introduces a phase
shift which 'splits' the beam on the sample \cite{nsplit}. Calling the
components of the split beam on the sample $\psi_{2}^{\pm}\left(
\overrightarrow{r_{s}}\right)  $ we have:%

\begin{align}
\psi_{2}^{\pm}\left(  \overrightarrow{r_{s}}\right)   &  =e^{ik_{o}y_{s}}\int
dk_{1}A\left(  k_{1}\right)  e^{ik_{1}y_{s}}e^{\pm ik_{1}\delta_{\text{NSE}%
}/2}e^{\pm i\omega_{z}L/v_{o}}\nonumber\\
&  =\Psi\,_{inc}\left(  \overrightarrow{r_{s}}\pm\delta_{\text{NSE}}/2\right)
e^{\pm i\omega_{z}L/v_{o}} \label{psi2}%
\end{align}
Since the measured quantity in spin echo $\left\langle \sigma_{x}\right\rangle
$ depends on the cross correlation between $\psi^{\pm}$ we calculate this
cross correlation for the wave functions (\ref{psi2})%

\begin{align}
\left\langle \psi_{2}^{+}\left(  \overrightarrow{r_{s}}\right)  \psi_{2}%
^{-*}\left(  \overrightarrow{r_{s}^{\prime}}\right)  \right\rangle  &
=\left\langle \Psi\,_{inc}\left(  \overrightarrow{r_{s}}+\delta_{\text{NSE}%
}/2\right)  \Psi\,_{inc}^{*}\left(  \overrightarrow{r_{s}^{\prime}}%
-\delta_{\text{NSE}}/2\right)  \right\rangle \nonumber\\
&  =R_{in-in}\left(  r_{s}-r_{s}^{\prime}+\delta_{\text{NSE}}\right)
=R_{in-in}\left(  \delta+\delta_{\text{NSE}}\right)
\end{align}
where $R_{in-in}\left(  \delta\right)  $ is the auto-correlation function of
the incident beam in the absence of NSE field.

\medskip Now following (\ref{two}) the wave function of neutrons scattered by
the sample that would reach the detector in the absence of a field in the
second coil is:%

\begin{equation}
\psi_{sc}^{\pm}\left(  \overrightarrow{r_{d}}\right)  =\frac{e^{ik_{2}%
r_{\text{d}}}}{r_{\text{d}}}\int d^{3}r_{s}e^{-i\overrightarrow{k_{2}}%
\cdot\overrightarrow{r_{s}}}\rho\left(  \overrightarrow{r_{s}}\right)
\psi_{2}^{\pm}\left(  \overrightarrow{r_{s}}\right)  \label{sed}%
\end{equation}
where $\left|  \overrightarrow{k_{2}}\right|  =\left|  \overrightarrow{k_{1}%
}\right|  =k_{o}$ (elastic scattering) and $\overrightarrow{k_{2}}$ is
directed along a line from the center of the sample to a point on the
detector, $\overrightarrow{r_{d}}$, so that $k_{o}r_{\text{d}}=\overrightarrow
{k_{2}}\cdot$ $\overrightarrow{r_{d}}$. Thus integrating over $d^{3}k_{2}$
will be equivalent to integrating over the detector area.

When the second NSE\ coil is turned on, the beam reaching the detector will
have an additional phase shift $\left(  e^{\mp i\omega_{\text{z}}L/v_{o}%
}e^{\mp ik_{2}\delta_{\text{NSE}}/2}\right)  $ so that at the detector we will have:%

\begin{equation}
\psi_{d}^{\pm}\left(  \overrightarrow{r_{d}}\right)  =e^{i\overrightarrow
{k_{2}}\cdot\overrightarrow{r_{d}}}\int d^{3}r_{s}e^{-i\overrightarrow{k_{2}%
}\cdot\overrightarrow{r_{s}}}e^{\mp ik_{2}\delta_{\text{NSE}}/2}\rho\left(
\overrightarrow{r_{s}}\right)  \Psi\,_{inc}\left(  \overrightarrow{r_{s}}%
\pm\delta_{\text{NSE}}/2\right)
\end{equation}
Then
\begin{align}
\left\langle \sigma_{x}\right\rangle  &  =\left\langle \psi_{d}\left(
\overrightarrow{r_{d}}\right)  \left|  \sigma_{x}\right|  \psi_{d}\left(
\overrightarrow{r_{d}}\right)  \right\rangle =\int d^{3}k_{2}\int d^{3}%
r_{s}\int d^{3}r_{s}^{\prime}e^{i\overrightarrow{k_{2}}\cdot\left(
\overrightarrow{r_{s}^{\prime}}-\overrightarrow{r_{s}}\right)  }%
e^{-ik_{2}\delta_{\text{NSE}}}\cdots\nonumber\\
&  \cdots\rho\left(  \overrightarrow{r_{s}}\right)  \Psi\,_{inc}\left(
\overrightarrow{r_{s}}+\delta_{\text{NSE}}/2\right)  \rho\left(
\overrightarrow{r_{s}^{\prime}}\right)  \Psi^{*}\,_{inc}\left(
\overrightarrow{r_{s}^{\prime}}-\delta_{\text{NSE}}/2\right) \nonumber
\end{align}
The integration over $d^{3}k_{2}$ gives $\delta\left(  \overrightarrow
{r_{s}^{\prime}}-\overrightarrow{r_{s}}-\delta_{\text{NSE}}\right)  $ so that
we have finally%

\begin{align}
\left\langle \sigma_{x}\right\rangle  &  =\int d^{3}r^{\prime}\left\langle
\rho\left(  \overrightarrow{r_{s}}\right)  \rho\left(  \overrightarrow{r_{s}%
}+\delta_{\text{NSE}}\right)  \right\rangle _{s}\left\langle \left|
\Psi\,_{inc}\left(  \overrightarrow{r_{s}}+\delta_{\text{NSE}}/2\right)
\right|  ^{2}\right\rangle _{b}\\
&  =G\left(  \delta_{\text{NSE}}\right)  R_{\text{in-in}}\left(  0\right)
=G\left(  \delta_{\text{NSE}}\right) \nonumber
\end{align}

Thus we see that in the case considered here with the range of available
$\overrightarrow{k_{2}}$ large enough, and the approximation that
$\delta_{\text{NSE}}$ is independent of $\overrightarrow{k_{2,1}}$ we get the
exact correlation function, independent of the properties of the incident
beam. Of course we know this from the classical treatment.

\section{Scattering from a time dependent system}

\smallskip In the present section we will show how the ideas presented in the
previous section can be applied to scattering from systems fluctuating in
time. In scattering from such systems the energy of the scattering radiation
is changed (inelastic or quasi-elastic scattering). For simplicity, in this
section we will neglect the position dependence of the scattering, considering
the scattering system to be concentrated at a single point, $\overrightarrow
{r_{s}}$ and restrict ourselves to massive particle scattering where the
energy is defined by choppers.\footnote{The use of crystals as energy defining
elements has been briefly discussed according to the present viewpoint in ref.
\cite{interlak} and will be discussed in more detail in a later work.} This is
equivalent to calculating the scattering integrated over all values of
$\overrightarrow{q\text{ }}$. We will treat the case of time and position
dependent scattering in the next section.

Under these restrictions equ. (\ref{one}) can be written:%

\begin{equation}
\psi_{sc}\left(  \overrightarrow{r_{d}},t_{d}\right)  =\int dt_{s}G_{o}\left(
\overrightarrow{r_{d}}-\overrightarrow{r_{s}},t_{d}-t_{s}\right)  V\left(
\overrightarrow{r_{s}},t_{s}\right)  \psi_{in}\left(  \overrightarrow{r_{s}%
},t_{s}\right)
\end{equation}
for the wave function at the detector located at $\overrightarrow{r_{d}}$ at
time $t_{d}$ where%

\begin{equation}
G_{o}\left(  \overrightarrow{r_{d}}-\overrightarrow{r_{s}},t_{d}-t_{s}\right)
=e^{i\frac{m\left|  \overrightarrow{r_{d}}-\overrightarrow{r_{s}}\right|
^{2}}{2\hbar\left(  t_{d}-t_{s}\right)  }} \label{greent}%
\end{equation}
is the Green's function (neglecting an unimportant prefactor) for the time
dependent unperturbed Schr\"{o}dinger equation.

The beam will be taken to be incident on the scatterer through a slit
(chopper) located at the origin that opens during the time interval $-T\leq
t_{o}\leq T$. (See fig. 6 for a definition of the notation used here.)
Analogous to the elastic case above, we consider the wave function incident on
the chopper to be completely uncorrelated for different values of $t_{o}$.
That is, we assume that the correlation time of the beam incident on the
chopper is much shorter than the correlation time that will be imposed on the
beam by the action of the chopper, or, in other words, the beam incident on
the chopper contains a much broader energy spectrum than will be selected by
the chopper system.

The wave function leaving the chopper and arriving at a point $\overrightarrow
{r_{s}}$ at time $t_{s}$ is given by \cite{mosh}, \cite{mf}, \cite{ggtdo}%

\begin{equation}
\psi_{in}\left(  \overrightarrow{r_{s}},t_{s}\right)  =\int_{-T}^{T}%
dt_{o}e^{i\frac{mr_{s}^{2}}{2\hbar\left(  t_{s}-t_{o}\right)  }}\psi
_{o}\left(  t_{o}\right)
\end{equation}
and%

\begin{align}
\left\langle \psi_{in}\left(  \overrightarrow{r_{s}},t_{s}\right)  \psi
_{in}^{*}\left(  \overrightarrow{r_{s}},t_{s}^{\prime}\right)  \right\rangle
&  =e^{-i\underline{\omega_{1}}\tau}e^{i2\underline{\omega_{1}}\xi_{s}%
\frac{\tau}{\underline{t}_{1}}}\int_{-T}^{T}dt_{o}e^{-i2\underline{\omega_{1}%
}\tau t_{o}/\underline{t}_{1}}\label{inint}\\
&  =e^{-i\underline{\omega_{1}}\tau}e^{i2\underline{\omega_{1}}\xi_{s}%
\frac{\tau}{\underline{t}_{1}}}2T\frac{\sin\alpha_{1}T}{\alpha_{1}T}\nonumber
\end{align}
where $\underline{\omega_{1}}=mr_{s}^{2}/2\hbar\underline{t_{1}^{2}}$

This emergence of plane wave-like states from the time dependent Green's
function has been discussed in \cite{feynhib}.

In deriving (\ref{inint}) we have taken%

\[
\left\langle \psi_{o}\left(  t_{o}\right)  \psi_{o}^{*}\left(  t_{o}^{\prime
}\right)  \right\rangle =\delta\left(  t_{o}-t_{o}^{\prime}\right)
\]

This (\ref{inint}) is the time analogue of the van Cittert-Zernike theorem
(\ref{rinsas}) for matter-wave optics.

Following arguments analogous to those in the previous section we find for the
intensity at the detector integrated over a time channel of width $2T_{d} $:%

\begin{equation}
\int\limits_{-T_{d}}^{T_{d}}dt_{d}\left|  \psi_{sc}\left(  \overrightarrow
{r_{d}},t_{d}\right)  \right|  ^{2}=\int\limits_{-T_{s}}^{T_{s}}d\xi_{s}\int
d\tau e^{i\omega\tau}e^{i\mu\xi_{s}}G_{s}\left(  \tau\right)  2T\frac
{\sin\alpha_{1}T}{\alpha_{1}T}2T_{d}\frac{\sin\alpha_{2}T_{d}}{\alpha_{2}%
T_{d}} \label{410}%
\end{equation}

\smallskip where%

\begin{equation}
\alpha_{1}=\frac{2\underline{\omega_{1}}\tau}{\underline{t_{1}}},\quad
\alpha_{2}=\frac{2\underline{\omega_{2}}\tau}{\underline{t_{2}}},\quad
\mu=\alpha_{1}+\alpha_{2} \label{alpha}%
\end{equation}

and $\omega=\underline{\omega_{2}}-\underline{\omega_{1}}$ is the mean energy
transfer of the scattered particles. There is a second chopper placed close to
the scatterer which opens for $\underline{t_{s}}-T_{s}\leq t_{s}\leq$
$\underline{t_{s}}+T_{s}$. The integral over $d\xi_{s}$ is seen to give
$2T_{s}\frac{\sin\mu T_{s}}{\mu T_{s}}$ with%

\begin{equation}
\mu=2\tau\left(  \frac{\underline{\omega_{2}}}{\underline{t_{2}}}%
+\frac{\underline{\omega_{1}}}{\underline{t_{1}}}\right)
\end{equation}
The term $e^{i\mu\xi_{s}}$ comes from the phase factor which is normally
neglected in optics. In the present case this factor has the effect that
moving the correlation interval $\left(  \tau\right)  $ (varying $\xi_{s}$) in
(\ref{410}) causes an additional difference in the two optical paths going
from the entrance chopper to the detector chopper via the two points separated
by $\tau$, \emph{i.e.} the paths \emph{1 }and \emph{2} in fig. 6). This
defines the influence of the sample chopper width $\left(  2T_{s}\right)  $ on
the resolution.

We then have:%

\begin{align}
\int_{-T_{d}}^{T_{d}}dt_{d}\left|  \psi_{sc}\left(  \overrightarrow{r_{d}%
},t_{d}\right)  \right|  ^{2}  &  =2T_{s}2T2T_{d}\int d\tau e^{i\omega\tau
}G_{s}\left(  \tau\right)  \mathbb{H}\label{td}\\
\mathbb{H}  &  =\frac{\sin\alpha_{2}T_{d}}{\alpha_{2}T_{d}}\frac{\sin\mu
T_{s}}{\mu T_{s}}\frac{\sin\alpha_{1}T}{\alpha_{1}T}%
\end{align}
where we have written $G_{s}\left(  \tau\right)  $ for $\left\langle
\rho\left(  \overrightarrow{r_{s}},t_{s}\right)  \rho^{*}\left(
\overrightarrow{r_{s}},t_{s}-\tau\right)  \right\rangle _{s}$ the time
dependent density-density correlation function of the scattering system.

Thus the scattering cross section is found to be proportional to the Fourier
transform of the density-density correlation function of the scattering system
multiplied by a function of $\tau$ which is the product of three functions
representing the effects of the opening times of the three choppers on the
resolution of the measurement. In this case the scattering is seen as the
interaction of the incoming state with the time fluctuations of the scattering
system. The wave function of the incoming state has an auto-correlation
function determined by the (time-dependent) optical properties of the defining
choppers. Only pairs of scattering events separated by times $\tau$, less than
the correlation time of the incoming wave contribute to the scattered wave and
their contribution is weighted by the beam auto-correlation function evaluated
at $\tau$.

\subsection{Discussion}

The situation represented by (\ref{td}) is completely analogous to the elastic
scattering case (\ref{qquutt}). The opening times $T_{1,2}$ influence the
resolution by an amount depending on $\underline{t}_{1,2}$ respectively, while
$T_{s}$ contributes to the uncertainty in the travel times in both arms and
depends on a weighted combination of the two travel times. The argument of
each term can be written as $\tau/\tau_{ci}$ where the $\tau_{ci}$ are
different correlation times. Clearly the measurement will be optimized when
all the $\tau_{ci}$ are approximately equal.

\section{A space and time dependent system}

Rewriting equ. (\ref{biggy}) we have for the intensity at the detector:%

\begin{equation}
\int d^{2}r_{d}dt_{d}\left|  \psi_{sc}\left(  \overrightarrow{r_{d}}%
,t_{d}\right)  \right|  ^{2}=\int d^{3}r_{s}dt_{s}\int d^{3}\delta d\tau
R_{out}\left(  \overrightarrow{\delta\text{ }},\tau,\overrightarrow{r_{2}%
},t_{2}\right)  G_{s}\left(  \overrightarrow{\delta\text{ }},\tau\right)
R_{in}\left(  \overrightarrow{\delta\text{ }},\tau,\overrightarrow{r_{1}%
},t_{1}\right)  \label{51}%
\end{equation}

with $G_{o}$ given by equ. \ref{greent}. We recall that%

\begin{equation}
\overrightarrow{r_{s}^{\prime}}=\overrightarrow{r_{s}}-\overrightarrow{\delta
},\quad t_{s}^{\prime}=t_{s}-\tau
\end{equation}

and we expand%

\begin{equation}
\left|  \overrightarrow{r_{s}^{\prime}}-\overrightarrow{r_{o}}\right|
^{2}\approx\left|  \overrightarrow{r_{s}}-\overrightarrow{r_{o}}\right|
^{2}-2\overrightarrow{\delta}\cdot\left(  \overrightarrow{r_{s}}%
-\overrightarrow{r_{o}}\right)
\end{equation}%

\begin{equation}
\frac{1}{t_{s}^{\prime}-t_{o}}\approx\frac{1}{t_{s}-t_{o}}+\frac{\tau}{\left(
t_{s}-t_{o}\right)  ^{2}}%
\end{equation}
where we have neglected terms of the order $\tau^{2}$ and $\delta^{2}$. To
calculate the product of the two Green's functions in (\ref{rin}) we need to
calculate the quantity%

\begin{align}
\Lambda_{in} &  =\frac{\left|  \overrightarrow{r_{s}}-\overrightarrow{r_{o}%
}\right|  ^{2}}{\left(  t_{s}-t_{o}\right)  }-\frac{\left|  \overrightarrow
{r_{s}^{\prime}}-\overrightarrow{r_{o}}\right|  ^{2}}{\left(  t_{s}^{\prime
}-t_{o}\right)  }=\\
&  =\frac{\left|  \overrightarrow{r_{s}}-\overrightarrow{r_{o}}\right|  ^{2}%
}{\left(  t_{s}-t_{o}\right)  }-\left(  \left|  \overrightarrow{r_{s}%
}-\overrightarrow{r_{o}}\right|  ^{2}-2\overrightarrow{\delta}\cdot\left(
\overrightarrow{r_{s}}-\overrightarrow{r_{o}}\right)  \right)  \times\left[
\frac{1}{t_{s}-t_{o}}+\frac{\tau}{\left(  t_{s}-t_{o}\right)  ^{2}}\right]  \\
&  =-\frac{\left|  \overrightarrow{r_{s}}-\overrightarrow{r_{o}}\right|
^{2}\tau}{\left(  t_{s}-t_{o}\right)  ^{2}}+\frac{2\overrightarrow{\delta
}\cdot\left(  \overrightarrow{r_{s}}-\overrightarrow{r_{o}}\right)  }{\left(
t_{s}-t_{o}\right)  }\label{lam}%
\end{align}

\smallskip Using our usual notation%

\begin{equation}
t_{s}\Rightarrow\underline{t}_{s}+\xi_{s}=\underline{t}_{1}+\xi_{s}%
\end{equation}%

\begin{equation}
\overrightarrow{r_{s}}\Rightarrow\overrightarrow{\underline{r}_{s}%
}+\overrightarrow{\epsilon_{s}}=\overrightarrow{\underline{r}}_{1}%
+\overrightarrow{\epsilon_{s}}%
\end{equation}

we have%

\begin{equation}
\Lambda_{in}=-\frac{\left|  \overrightarrow{r_{s}}-\overrightarrow{r_{o}%
}\right|  ^{2}\tau}{\underline{t}_{1}^{2}}\left(  1-2\frac{\xi_{s}-t_{o}%
}{\underline{t}_{1}}\right)  +\frac{2\overrightarrow{\delta}\cdot\left(
\overrightarrow{r_{s}}-\overrightarrow{r_{o}}\right)  }{\underline{t}_{1}%
}\left(  1-\frac{\xi_{s}-t_{o}}{\underline{t}_{1}}\right)
\end{equation}

and finally%

\begin{align}
\Lambda_{in}  &  =-\frac{\left|  \overrightarrow{\underline{r}}_{1}\right|
^{2}\tau}{\underline{t}_{1}^{2}}+\frac{2\overrightarrow{\delta}\cdot\left(
\overrightarrow{\underline{r}}_{1}\right)  }{\underline{t}_{1}}+t_{o}\left(
-\frac{2\left|  \overrightarrow{\underline{r}}_{1}\right|  ^{2}\tau
}{\underline{t}_{1}^{3}}+\frac{2\overrightarrow{\delta}\cdot\overrightarrow
{\underline{r}}_{1}}{\underline{t}_{1}^{2}}\right)  +\overrightarrow{r_{o}%
}\cdot\left(  \frac{2\overrightarrow{\underline{r}}_{1}\tau}{\underline{t}%
_{1}^{2}}-\frac{2\overrightarrow{\delta}}{\underline{t}_{1}}\right)
\label{lamin}\\
&  +\xi_{s}\left(  -\frac{2\overrightarrow{\delta}\cdot\left(  \overrightarrow
{\underline{r}}_{1}\right)  }{\underline{t}_{1}^{2}}+\frac{2\left|
\overrightarrow{\underline{r}}_{1}\right|  ^{2}\tau}{\underline{t}_{1}^{3}%
}\right)  +\overrightarrow{\epsilon_{s}}\cdot\left(  \frac{2\overrightarrow
{\delta}}{\underline{t}_{1}}-\frac{2\overrightarrow{\underline{r}}_{1}\tau
}{\underline{t}_{1}^{2}}\right)
\end{align}

Thus the phase of the integrand of $R_{in}$ (equ. \ref{rin}) is:%

\begin{equation}
\frac{m}{2\hbar}\Lambda_{in}=-\underline{\omega}_{1}\tau+\underline
{\overrightarrow{k}}_{1}\cdot\overrightarrow{\delta}+\left(  t_{o}-\xi
_{s}\right)  \alpha_{1}+\left(  \overrightarrow{r_{o}}-\overrightarrow
{\epsilon_{s}}\right)  \cdot\overrightarrow{\beta_{1}}%
\end{equation}
where%

\begin{equation}
\underline{\omega}_{1}=\frac{m}{2\hbar}\frac{\left|  \overrightarrow
{\underline{r}}_{1}\right|  ^{2}}{\underline{t}_{1}^{2}},\quad\underline
{\overrightarrow{k}}_{1}=\frac{m}{\hbar}\frac{\overrightarrow{\underline{r}%
}_{1}}{\underline{t}_{1}}%
\end{equation}%

\begin{equation}
\alpha_{1}=\left(  -\frac{2\underline{\omega}_{1}\tau}{\underline{t}_{1}%
}+\frac{\overrightarrow{\delta}\cdot\underline{\overrightarrow{k}}_{1}%
}{\left(  \underline{t}_{1}\right)  }\right)  ,\quad\overrightarrow{\beta_{1}%
}=\left(  \frac{\overrightarrow{\underline{k}}_{1}\tau}{\underline{t}_{1}%
}-\frac{m\overrightarrow{\delta}}{\hbar\left(  \underline{t}_{1}\right)
}\right)  \label{alphab}%
\end{equation}

Then%

\begin{equation}
R_{in}=\int\limits_{y=-a}^{y=a}d^{2}r_{o}\int\limits_{-T}^{T}dt_{o}%
e^{i\frac{m}{2\hbar}\Lambda_{in}}=e^{i\left(  -\underline{\omega}_{1}%
\tau+\underline{\overrightarrow{k}}_{1}\cdot\overrightarrow{\delta}\right)
}e^{-i\left(  \xi_{s}\alpha_{1}+\overrightarrow{\epsilon_{s}}\cdot
\overrightarrow{\beta}_{1}\right)  }2al\frac{\sin\beta_{1y}a}{\beta_{1y}%
a}2T\frac{\sin\alpha_{1}T}{\alpha_{1}T} \label{rinin}%
\end{equation}
where we have specialized to the case of a one dimensional input slit going
from $y=-a$ to $y=a$. $l(\gg a)$ is the height of the slit. Note that
$\overrightarrow{\underline{k}}_{1}$ is defined to be parallel to the $x$ axis
so that $\left(  \overrightarrow{\underline{k}}_{1}\right)  _{y}=0$. This
result (\ref{rinin}) is the generalization of the van Cittert-Zernike theorem
to the case of time dependent matter wave optics. Comparison with the pure
spatial case (\ref{rinsas}) and the pure time dependent case (\ref{inint})
shows that these are special limits of the general case (\ref{rinin}).

By our symmetry rules (\ref{symrule}) $R_{out}$ can be written%

\begin{equation}
\int\limits_{Det}d^{2}r_{d}\int_{-T_{d}}^{T_{d}}dt_{d}e^{i\frac{m}{2\hbar
}\Lambda_{out}}=e^{i\left(  \underline{\omega}_{2}\tau-\underline
{\overrightarrow{k}}_{2}\cdot\overrightarrow{\delta}\right)  }e^{i\left(
\xi_{s}\alpha_{2}-\overrightarrow{\epsilon_{s}}\cdot\overrightarrow{\beta}%
_{2}\right)  }2dl\frac{\sin\beta_{2y^{\prime}}d}{\beta_{2y^{\prime}}d}%
2T_{d}\frac{\sin\alpha_{2}T_{d}}{\alpha_{2}T_{d}}%
\end{equation}
where $y^{\prime}$ indicates the direction perpendicular to the output arm
$\left(  \overrightarrow{\underline{r}}_{2}=\overrightarrow{\underline{r}}%
_{d}-\overrightarrow{\underline{r}}_{s}\right)  $ and%

\begin{equation}
\alpha_{2}=\left(  \frac{2\underline{\omega}_{2}\tau}{\underline{t}_{2}}%
-\frac{\overrightarrow{\delta}\cdot\underline{\overrightarrow{k}}_{2}%
}{\underline{t}_{2}}\right)  ,\quad\overrightarrow{\beta_{2}}=\left(
-\frac{\overrightarrow{\underline{k}}_{2}\tau}{\underline{t}_{2}}%
+\frac{m\overrightarrow{\delta}}{\hbar\underline{t}_{2}}\right)
\end{equation}

Performing the integral in (\ref{51}) over the sample volume and scattering
time yields%

\begin{equation}
\int d^{3}\epsilon_{s}\int_{-T_{s}}^{T_{s}}d\xi_{s}\cdots=2T_{s}\frac{\sin\mu
T_{s}}{\mu T_{s}}l\int_{0}^{R_{s}}\epsilon_{s}d\epsilon_{s}\int_{0}^{2\pi
}d\theta e^{i\epsilon_{s}\nu\cos\theta} \label{ints}%
\end{equation}

\smallskip where we assumed the sample to be in the shape of a cylinder of
radius $R_{s}$. In equ. (\ref{ints})%

\begin{equation}
\mu=\alpha_{2}-\alpha_{1}=\left(  -\overrightarrow{\delta}\cdot\left(
\frac{\overrightarrow{\underline{k}}_{1}}{\underline{t}_{1}}+\frac
{\overrightarrow{\underline{k}}_{2}}{\underline{t}_{2}}\right)  +2\tau\left(
\frac{\underline{\omega}_{1}}{\underline{t}_{1}}+\frac{\underline{\omega}_{2}%
}{\underline{t}_{2}}\right)  \right)
\end{equation}
and%

\begin{equation}
\nu=\left|  \overrightarrow{\beta}_{2}-\overrightarrow{\beta}_{1}\right|
=\left|  \frac{m\overrightarrow{\delta}}{\hbar}\left(  \frac{1}{\underline
{t}_{1}}+\frac{1}{\underline{t}_{2}}\right)  -\tau\left(  \frac
{\overrightarrow{\underline{k}}_{1}}{\underline{t}_{1}}+\frac{\overrightarrow
{\underline{k}}_{2}}{\underline{t}_{2}}\right)  \right|
\end{equation}

The two dimensional integral gives%

\begin{equation}
2\pi\int_{0}^{R_{s}}\epsilon_{s}d\epsilon_{s}J_{0}\left(  \nu\epsilon
_{s}\right)  =2\pi R_{s}^{2}\frac{J_{1}\left(  \nu R_{s}\right)  }{\nu R_{s}}%
\end{equation}
The result of evaluating (\ref{51}) is then%

\begin{equation}
\int d^{3}r_{d}dt_{d}\left|  \psi_{sc}\left(  \overrightarrow{r_{d}}%
,t_{d}\right)  \right|  ^{2}=\int d^{3}\delta d\tau e^{i\left(  \omega
\tau-\overrightarrow{q}\cdot\overrightarrow{\delta}\right)  }G_{s}\left(
\overrightarrow{\delta},\tau\right)  \mathbb{H} \label{526}%
\end{equation}
where%

\begin{equation}
\omega=\underline{\omega}_{2}-\underline{\omega}_{1},\quad\overrightarrow
{q}=\overrightarrow{\underline{k}}_{2}-\overrightarrow{\underline{k}}_{1}%
\end{equation}%

\begin{equation}
\mathbb{H}=2^{5}\left(  d\frac{\sin\beta_{2y^{\prime}}d}{\beta_{2y^{\prime}}%
d}\right)  \left(  T_{d}\frac{\sin\alpha_{2}T_{d}}{\alpha_{2}T_{d}}\right)
\left(  T_{s}\frac{\sin\mu T_{s}}{\mu T_{s}}\right)  \left(  2\pi R_{s}%
^{2}\frac{J_{1}\left(  \nu R_{s}\right)  }{\nu R_{s}}\right)  \left(
T_{o}\frac{\sin\alpha_{1}T_{o}}{\alpha_{1}T_{o}}\right)  \left(  a\frac
{\sin\beta_{1y}a}{\beta_{1y}a}\right)  \label{hehe}%
\end{equation}

\subsection{Discussion}

For convenience we rewrite below%

\[
\alpha_{1}=\frac{\underline{\overrightarrow{k}}_{1}}{\underline{t}_{1}}%
\cdot\left(  \overrightarrow{\delta}-\overrightarrow{v}_{1}\tau\right)
,\quad\overrightarrow{\beta_{1}}=-\frac{m}{\hbar\underline{t}_{1}}\left(
\overrightarrow{\delta}-\overrightarrow{v}_{1}\tau\right)  \quad
\]%

\begin{equation}
\alpha_{2}=-\frac{\underline{\overrightarrow{k}}_{2}}{\underline{t}_{2}}%
\cdot\left(  \overrightarrow{\delta}-\overrightarrow{v}_{2}\tau\right)
,\quad\overrightarrow{\beta_{2}}=\frac{m}{\hbar\underline{t}_{2}}\left(
\overrightarrow{\delta}-\overrightarrow{v}_{2}\tau\right)  \quad\label{528}%
\end{equation}%

\[
\mu=\left(  \alpha_{2}-\alpha_{1}\right)  ,\quad\nu=\left|  \overrightarrow
{\beta}_{2}-\overrightarrow{\beta}_{1}\right|
\]

$\alpha_{1}$ and $\beta_{1}$ represent the influence of the input slit and
chopper on the overall resolution, $\mu$ and $\nu$ the influence of the sample
size and chopper while $\alpha_{2}$ and $\beta_{2}$ represent the effects of
the detector slit and time resolution.

Each of these variables has a term proportional to $\tau$ and a term
proportional to $\delta$. Keeping in mind the Fourier transform (\ref{526}) we
see that the terms proportional to $\tau$ will give the energy resolution and
those proportional to $\delta$ will give the $q$ resolution. Thus, for
example, the $\tau$ term in $\alpha_{1}$ represents the effects of the first
chopper opening time, $T_{o}$, on the energy resolution just as in the pure
time dependent case, equations (\ref{alpha}, \ref{td}) while the $\delta$ term
represents the effect of the first chopper width on the $q$ resolution,
\emph{i.e.} the chopper width results in an uncertainty in $\overrightarrow
{k}_{1}$ that gives a contribution to the $q$ resolution. Similar remarks hold
for each of the terms in (\ref{528}). Note that by definition $\underline
{\overrightarrow{k_{1}}}_{y},\underline{\overrightarrow{k_{2}}}_{y^{\prime}%
}=0$ but we display these terms in (\ref{528}) to keep the symmetry.

We see that our approach yields directly the mutual influence of the $q$ and
$\omega$ resolutions in a compact analytic form. Since each element of the
apparatus is associated with a definite factor in $\mathbb{H}$, we see
directly the influence of the individual elements on the overall resolution.
Modification of these elements can readily be accommodated as shown in the
next section.

If we wish we can follow the more usual procedure of replacing each term by a
Gaussian function with the appropriate width. This then allows an easy way to
estimate the overall width of the $\omega,q$ resolution (given by the
convolution of the Fourier transform of each term in $\mathbb{H}$) as the
square root of the sum of the squares of the widths of each term and shows
that the optimum is when all the widths are equal.

Each factor in (\ref{hehe}) is associated with a certain velocity in $\left(
\overrightarrow{\delta},\tau\right)  $ space so that the relevant correlation
volumes can be considered as moving as a function of $\tau$. For example the
volume associated with $\alpha_{2}$ is moving with a velocity $\overrightarrow
{\underline{v}}_{2}$, the nominal velocity of the scattered wave. Thus we see
that the coherence volume associated with the detector parameters can be
considered as consisting of the region limited by $\pm\hbar\underline{t}%
_{2}/md$ in the $\delta_{y}^{\prime}$ direction and $\pm\underline{t}%
_{2}/T_{d}k_{2}$ in the $\delta_{x^{\prime}}$ direction, moving (as function
of $\tau$) with velocity $v_{2}$ in the $x^{\prime}$ direction (direction of
the scattered beam). Each of the other pairs of terms contributes a similar
moving limit to the region of $\left(  \overrightarrow{\delta},\tau\right)  $ space.

In a forthcoming work we will present a detailed discussion of the shapes and
behavior of these correlation volumes for a series of instruments.

\subsection{Entrance slit with moving edges}

We now modify the above treatment to consider the more realistic case where
the chopper slits move with finite speed. See \cite{ggtdo} for discussion of a
related problem. We consider that the slit width depends on the time $t_{o} $,
at which the beam passes through the slit in the following manner:%

\begin{equation}
a\left(  t_{o}\right)  =\left\{
\begin{array}
[c]{lll}%
0 & if & t_{o}<-T_{1}\\
a+\theta t_{o} & if & -T_{1}\leq t_{o}\leq0\\
a-\theta t_{o} & if & 0<t_{o}\leq T_{1}\\
0 & if & T_{1}<t_{o}%
\end{array}
\right.
\end{equation}
where $a$ is the maximum opening of the slit (which occurs at $t_{o}=0$) and
$\theta=a/T_{1}$. Then we rewrite equ. (\ref{rinin}) as%

\begin{align}
R_{in}  &  =e^{i\Gamma}l\int_{-T_{o}}^{T_{o}}dt_{o}e^{i\alpha t_{o}}%
\int_{-a\left(  t_{o}\right)  }^{a\left(  t_{o}\right)  }dye^{i\beta_{1y}y}\\
&  =e^{i\Gamma}l\int_{-T_{o}}^{T_{o}}dt_{o}e^{i\alpha t_{o}}\frac{\left[
e^{i\beta_{1y}a\left(  t_{o}\right)  }-e^{-i\beta_{1y}a\left(  t_{o}\right)
}\right]  }{i\beta_{1y}}%
\end{align}

\smallskip where $\Gamma$ represents those terms in equ. (\ref{lamin}) which
do not depend on $t_{o}$ and $r_{o}$.

The last integral can be evaluated to give%

\begin{equation}
R_{in}=e^{i\Gamma}l\left[  2aT_{o}\frac{\sin\frac{1}{2}\left(  \alpha_{1}%
T_{o}-\beta_{1y}a\right)  }{\frac{1}{2}\left(  \alpha_{1}T_{o}-\beta
_{1y}a\right)  }\frac{\sin\frac{1}{2}\left(  \alpha_{1}T_{o}+\beta
_{1y}a\right)  }{\frac{1}{2}\left(  \alpha_{1}T_{o}+\beta_{1y}a\right)
}\right]  \label{rnew}%
\end{equation}
The rest of the calculation proceeds as above so the only change in the result
is to replace the $\sin x/x$ functions depending on $a,T_{o}$ and their
associated prefactors in (\ref{hehe}) with the function in square brackets in
(\ref{rnew}). The correlation function (\ref{rnew}) is seen to have the
correct normalization at $\alpha_{1}=\beta_{1y}=0$, \emph{i.e.} at
$\overrightarrow{\delta}=\tau=0$. The two terms have separated maxima
occurring at%

\begin{equation}
\delta_{x}\pm\delta_{y}\frac{a}{\underline{v}_{1}T_{o}}-\underline{v}_{1}%
\tau=0
\end{equation}
where $\delta_{x}$ is the component of $\overrightarrow{\delta}$ along the
direction of the incident beam and $\delta_{y}$ its component in the
perpendicular direction. Thus for $\tau=0$ the maxima lie on the lines through
the origin:%

\begin{equation}
\frac{\delta_{y}}{\delta_{x}}=\pm\frac{\underline{v}_{1}T_{o}}{a}=\pm
\frac{\underline{v}_{1}}{v_{s}}%
\end{equation}
where $v_{s}$ is the velocity of the slit boundary. For non-zero $\tau$ the
slope of the lines stays the same but the crossing point moves along the
$\delta_{x}$ axis with a velocity $\underline{v}_{1}$. The correlation volume
for the input beam, \emph{i.e. }the overlap region where the two factors in
(\ref{rnew}) are significant, is thus seen to be given by a complex interplay
between the slopes of the lines and the widths of the $\sin x/x$ functions
which depend on the correlation lengths $\delta_{yc}=L_{1}/\underline{k}%
_{1}a,\ \delta_{xc}=\underline{t}_{1}/T_{o}\underline{k}_{1},$ and the
correlation time $\tau=\underline{t}_{1}/2\underline{\omega}_{1}T_{o}$.

This is a demonstration of some of the advantages of the present method,
showing how a change of a component can be easily accommodated and how complex
relationships between the energy and momentum resolution can be represented in
a straight forward manner.

\section{Conclusions}

Starting with Fermi's 'golden rule' for perturbation theory and assuming the
incident and final wave functions of the scattered particles to be plane
waves, van Hove \cite{vanh} showed that the scattering cross section is
proportional to a function of the momentum and energy transfer, $S\left(
q,\omega\right)  $, which in turn is the Fourier transform of the time and
space dependent auto-correlation function of the density fluctuations of the
scattering system.

However since no experiment is ever done with perfect resolution, measurements
yield the convolution of $S\left(  q,\omega\right)  $ with an instrument
resolution function. The determination of the resolution and its use in the
interpretation of the data are left to the experimenters.

In this conventional approach a scattering experiment is divided up into two
separate parts which are treated by two, different, mutually exclusive
approximations. The scattering event in the sample is treated by assuming
infinitely extended plane waves for the incident and scattered beams, while
the beam is treated as consisting of classical point particles in order to
describe its motion through the apparatus. This same, dual approach is also
used to discuss multiple scattering: the scattering events are described as a
scattering of plane wave states while the motion between two scattering points
is treated as a classical trajectory. We hope to present a unified treatment
of multiple scattering using the present approach in the near future.

In the present work, the entire scattering experiment is described as a
wave-optical phenomenon. This more general approach provides a rigorous
justification for the usual dual approximation since, in most cases the
correlation volumes are much smaller than the geometrical dimensions (beam
cross sections, chopper pulse lengths, etc.) involved. However in some cases
this is not true and the complete wave-optical description of the global
system of instrument and scattering sample must be used. In addition to
precision light scattering experiments such cases include perfect crystal
neutron optics (\emph{e.g. }interferometry) where the definition of the wave
vector is so precise that the correlation lengths can be comparable to the
geometrical dimensions. These cases are sometimes referred to as ''spherical
wave effects''\cite{SHULL}. In fact these effects are natural consequences of
the full wave-optical analysis presented here.

Usual methods of determining the resolution include measurement of a known
calibration sample, Monte Carlo calculations and estimates based on assuming a
Gaussian form for the separate contributions to the overall resolution so that
the separate contributions can be combined using the sum of squares rule.

One method of carrying out the deconvolution is to divide the Fourier
transform of the measured data by the Fourier transform of the resolution
function. Usually however a model form is assumed for $S\left(
\overrightarrow{q},\omega\right)  $ and some parameters of the model are
varied until the convolution with the resolution function gives agreement with
the measured data.

In the present work we calculate the incoming wave function as it is shaped by
transmission through the elements of the apparatus according to the laws of
optics. For example on propagating through a slit and/or a chopper a wave
acquires a correlation function with a finite width. After scattering, the
individual waves scattered from a neighboring pair of points in the sample
lose correlation on passing through the post-sample elements of the apparatus.
We have shown that the probability of a scattered particle arriving at a
detector at a given space-time location is proportional to the Fourier
transform of the density fluctuation auto-correlation function (as shown by
van Hove) but multiplied by a function $\mathbb{H}\left(  \overrightarrow
{\delta},\tau\right)  $ which represents the resolution of the measurement and
is the Fourier transform of the instrument resolution as it is normally
considered. This function is the product of a series of functions, each
representing the effect of a single element of the apparatus on the
measurement. Changes in one element can be easily accommodated by changes in
the appropriate function. The regions of $\left(  \overrightarrow{\delta}%
,\tau\right)  $ space where each function is significant are the correlation
volumes for the different elements of the apparatus. A measurement can be
optimized by maximizing the region of overlap of all the correlation volumes.
The fact that these volumes are seen to be moving as a function of the
correlation time $\tau$, is a reflection of the fact that the $\omega$, and
$\overrightarrow{q}$ resolutions depend on each other to some extent and we
have an analytic description of this dependence. The present analysis offers a
graphical method of optimizing each measurement, with the influence of all
parameters such as the scattering angle included. In a forthcoming work we
will give detailed examples of how this can be carried out for various types
of scattering measurements.

\smallskip

\section{Appendix - Normalization and calculation of cross sections}

In the main text we have neglected all prefactors and normalizing constants in
the Green's functions and incident wave function. In this appendix we will
show how inclusion of these factors leads to the usual expression for the
cross section and give a proper calculation of $\psi_{in}$ as mentioned in
footnote 2 of section 2. We will treat the most general case of space and time
dependent density fluctuations (sec. 5).

\subsection{Normalization of the Green's functions}

\subsubsection{Calculation of the incident wave.}

To calculate the wave function incident on the scattering sample we use%

\begin{equation}
\psi_{in}\left(  \overrightarrow{x},t\right)  =\frac{-1}{4\pi}\int_{0}^{t_{+}%
}dt_{o}\int d^{2}S_{o}\psi_{o}\left(  \overrightarrow{x}_{o},t_{o}\right)
\overrightarrow{n}\cdot\overrightarrow{\nabla_{o}}\widetilde{G}\left(
\overrightarrow{x}-\overrightarrow{x}_{o},t-t_{o}\right)  \label{qpsin}%
\end{equation}

where the integral over $d^{2}S_{o}$ is taken over the plane of the entrance
slit, $z_{o}=0,$ and $\psi_{o}$ is non-zero only in the entrance slit $\left(
-a<y_{o}<a\right)  $. (We make the Kirchhoff approximation as is usual in
optics). The function $\widetilde{G}$ satisfies the boundary condition
$\widetilde{G}\left(  z_{o}=0\right)  =0$ and can be taken as%

\begin{equation}
\widetilde{G}=g\left(  \overrightarrow{r}_{+},t\right)  -g\left(
\overrightarrow{r}_{-},t\right)
\end{equation}
where%

\begin{equation}
\left(  \overrightarrow{r}_{\pm}\right)  ^{2}=\left(  x-x_{o}^{2}\right)
+\left(  y-y_{o}\right)  ^{2}+\left(  z\mp z_{o}\right)  ^{2}%
\end{equation}
and $g(\overrightarrow{r},t)$ satisfies%

\begin{equation}
\nabla^{2}g-\frac{2m}{i\hbar}\frac{\partial g}{\partial t}=-4\pi\delta
^{(3)}\left(  \overrightarrow{r}\right)  \delta(t)
\end{equation}
$g(\overrightarrow{r},t)$ is then normalized as \cite{mf}:%

\begin{equation}
g(\overrightarrow{r},t)=\left(  \frac{2\pi i\hbar}{m}\right)  \left(  \frac
{m}{2\pi i\hbar t}\right)  ^{3/2}e^{i\frac{mr^{2}}{2\hbar t}}%
\end{equation}
Then (equ. \ref{qpsin})%

\begin{equation}
\psi_{in}\left(  \overrightarrow{x},t\right)  =z\left(  \frac{m}{2\pi i\hbar
}\right)  ^{3/2}\int_{0}^{t_{+}}\frac{dt_{o}}{\left(  t-t_{o}\right)  ^{5/2}%
}\int d^{2}S_{o}\psi_{o}\left(  \overrightarrow{x}_{o},t_{o}\right)
\underline{G}\left(  \overrightarrow{x}-\overrightarrow{x}_{o},t-t_{o}\right)
\end{equation}
where $\underline{G}=$ $\exp\left(  i\frac{mr^{2}}{2\hbar t}\right)  $is the
Green's function without prefactor as used in the main text.

\subsubsection{Calculation of the scattered wave.}

Taking the interaction potential as%

\begin{equation}
V\left(  \overrightarrow{r},t\right)  =2\pi\frac{\hbar^{2}b}{m}\rho\left(
\overrightarrow{r},t\right)
\end{equation}
where $b$ is the scattering length and $\rho$ is the number density of
scatterers the scattered wave is found to be (making the Born approximation)%

\begin{equation}
\psi_{sc}\left(  \overrightarrow{r},t\right)  =\frac{2\pi\hbar b}{im}\int
G_{o}\left(  \overrightarrow{r}-\overrightarrow{r}_{s},t-t_{s}\right)
\rho\left(  \overrightarrow{r}_{s},t_{s}\right)  \Psi_{in}\left(
\overrightarrow{r}_{s},t_{s}\right)  d^{3}r_{s}dt_{s} \label{qpsisc}%
\end{equation}
where $G_{o}$ satisfies%

\begin{equation}
i\hbar\frac{\partial G_{o}}{\partial t}+\frac{\hbar^{2}}{2m}\nabla^{2}%
G_{o}=\frac{-\hbar}{i}\delta^{(3)}\left(  \overrightarrow{r}\right)  \delta(t)
\end{equation}
and is normalized as \cite{feynhib}%

\begin{equation}
G_{o}=\left(  \frac{m}{2\pi i\hbar t}\right)  ^{3/2}\underline{G}%
\end{equation}
so that (\ref{qpsisc})%

\begin{equation}
\psi_{sc}\left(  \overrightarrow{r},t\right)  =\left(  \frac{2\pi\hbar b}%
{im}\right)  \left(  \frac{m}{2\pi i\hbar}\right)  ^{3}\frac{z_{1}}%
{\underline{t}_{2}^{3/2}\underline{t}_{1}^{5/2}}\int\int dt_{o}d^{2}S_{o}%
\int\int\underline{G}\rho\underline{G}\psi_{o}d^{3}r_{s}dt_{s}%
\end{equation}
where $\underline{t}_{1,2}$ are the average flight times in the two arms of
the apparatus and $z_{1}$ is the distance from the entrance slit to the center
of the sample. We have neglected the variation of the prefactors over the
regions of integration as these are small compared to the distances and flight
times involved. With suitable normalization of the incoming wave,
$\psi_{o\text{,}}$ the intensity, $I_{d}$, counted in the detector during the
time interval $2T_{d}$, will be given by%

\begin{equation}
v_{2}\int d^{2}r_{d}\int\limits_{-T_{d}}^{T_{d}}dt_{d}\left|  \psi
_{sc}\right|  ^{2}=v_{2}b^{2}\left(  \frac{m}{2\pi\hbar}\right)  ^{4}%
\Sigma_{o}\frac{z_{1}^{2}}{\underline{t}_{2}^{3}\underline{t}_{1}^{5}}%
l^{2}\int\int d^{3}r_{s}dt_{s}\int\int d^{3}\delta d\tau R_{out}\left\langle
\rho\rho^{\prime}\right\rangle R_{in} \label{qint1}%
\end{equation}
where $R_{out,in}$ are defined as in section 2 and $\left\langle \rho
\rho^{\prime}\right\rangle $ is related to the van Hove correlation function
$G_{s}\left(  \overrightarrow{\delta},\tau\right)  $ by \cite{vanh}%

\begin{equation}
\left\langle \rho\rho^{\prime}\right\rangle =\rho_{s}G_{s}\left(
\overrightarrow{\delta},\tau\right)
\end{equation}
and the correlation function of the input beam is defined to be%

\begin{equation}
\left\langle \psi_{o}\left(  \overrightarrow{r},t\right)  \psi_{o}^{*}\left(
\overrightarrow{r}^{\prime},t^{\prime}\right)  \right\rangle _{inc}=\Sigma
_{o}\delta^{(2)}\left(  \overrightarrow{r}-\overrightarrow{r}^{\prime}\right)
\delta\left(  t-t^{\prime}\right)  \label{qcorr1}%
\end{equation}

$l$ is the height of the beam perpendicular to the scattering plane.

The multiple integral in (\ref{qint1}) has been evaluated in (\ref{526}).
Setting all the slit widths and chopper opening times to zero in the function
$\mathbb{H}$ so that this takes on its maximum value (note: $\lim
_{x\rightarrow0}J_{1}\left(  x\right)  /x=1/2$) we get from (\ref{qint1}) for
the intensity $I_{d}$%

\begin{equation}
I_{d}=v_{2}b^{2}\left(  \frac{m}{2\pi\hbar}\right)  ^{4}\Sigma_{o}\frac
{z_{1}^{2}}{\underline{t}_{2}^{3}\underline{t}_{1}^{5}}l^{3}\mathbb{K}\left(
\pi R_{s}^{2}\right)  \rho_{s}\frac{2\pi}{N}S\left(  \overrightarrow{q}%
,\omega\right)
\end{equation}
where we have used van Hove's definition \cite{vanh}:%

\begin{equation}
S\left(  \overrightarrow{q},\omega\right)  =\frac{N}{2\pi}\int\int d^{3}\delta
d\tau e^{i\left(  \overrightarrow{q}\cdot\overrightarrow{\delta}-\omega
\tau\right)  }G_{s}\left(  \overrightarrow{\delta},\tau\right)
\end{equation}
$N=\rho_{s}\left(  \pi R_{s}^{2}l\right)  $ is the total number of scattering
centers in the sample and%

\begin{equation}
\mathbb{K=}\left(  2T_{o}\right)  \left(  2a\right)  \left(  2T_{s}\right)
\left(  2T_{d}\right)  \left(  2d\right)
\end{equation}
Since $z_{1}=v_{1}\underline{t}_{1}\left(  =L_{1}\right)  $ we finally obtain:%

\begin{equation}
I_{d}=v_{2}b^{2}\left(  \frac{m}{\hbar}\right)  ^{4}\frac{\Sigma_{o}}{\left(
2\pi\right)  ^{3}}\frac{v_{1}^{2}}{\underline{t}_{2}^{3}\underline{t}_{1}^{3}%
}l^{2}\mathbb{K}S\left(  \overrightarrow{q},\omega\right)  \label{qint2}%
\end{equation}

\subsection{Calculation of the cross section}

\subsubsection{Normalization of the input wave correlation function}

Rewriting the definition (\ref{qcorr1}) we have%

\begin{align}
\left\langle \psi_{o}\psi_{o}^{*\prime}\right\rangle _{inc}  &  =\Sigma
_{o}\delta^{(2)}\left(  \overrightarrow{r}-\overrightarrow{r}^{\prime}\right)
\delta\left(  t-t^{\prime}\right)  =\nonumber\\
&  =\Sigma_{o}\frac{1}{\left(  2\pi\right)  ^{2}}\int d^{2}%
ke^{i\overrightarrow{k}\cdot\left(  \overrightarrow{r}-\overrightarrow
{r}^{\prime}\right)  }\frac{1}{2\pi}\int d\omega e^{-i\omega\left(
t-t^{\prime}\right)  }%
\end{align}
so that%

\begin{equation}
\left\langle \rho\right\rangle =\left\langle \left|  \psi\right|
^{2}\right\rangle =\left\langle \psi_{o}\psi_{o}^{*\prime}\right\rangle
_{\delta=\tau=0}=\frac{\Sigma_{o}}{\left(  2\pi\right)  ^{3}}\int\int
d^{2}kd\omega\rightarrow\infty
\end{equation}
for the white spectrum we are assuming. From this we conclude that the density
fluctuations contained in the interval $d^{2}kd\omega$ are given by%

\begin{equation}
\frac{d^{3}\rho}{d\omega dk^{2}}d^{2}kd\omega=\frac{\Sigma_{o}}{\left(
2\pi\right)  ^{3}}d^{2}kd\omega
\end{equation}
or%

\begin{equation}
\frac{d^{3}\rho}{d\omega dk^{2}}=\frac{\Sigma_{o}}{\left(  2\pi\right)  ^{3}}%
\end{equation}

\subsubsection{Calculation of the intensity in terms of the cross section.}

The incident flux in a velocity interval $dv_{1}$ can be written%

\begin{equation}
v_{1}\left(  \frac{d^{3}\rho}{d\omega dk^{2}}\right)  \left(  \frac{d\omega
}{dv_{1}}\right)  dv_{1}k_{1}^{2}d\Omega_{1}%
\end{equation}
with $v_{1}$ the incident velocity and $d\Omega_{1}$ the input solid angle.
Multiplying by the chopper opening time $\left(  2T_{o}\right)  $ and the area
of the input slit $\left(  2al\right)  $ we obtain the total number of
particles contained in one pulse. The fraction of particles scattered into the
velocity range $dv_{2}$ and solid angle $d\Omega_{2}$ is then given by%

\begin{equation}
\left(  \frac{1}{A_{s}}\right)  \frac{d^{2}\sigma}{d\omega d\Omega}%
\frac{d\omega}{dv_{2}}dv_{2}d\Omega_{2}%
\end{equation}
where in the first factor $A_{s}$ is the area of the sample exposed to the
beam. We use the cross section definition of van Hove which applies to the
whole sample (\emph{i.e.} $N$ times the cross section for the individual
scattering atoms). Now with%

\begin{equation}
\frac{d\omega}{dv}=\frac{mv}{\hbar},\quad dv_{1,2}=\frac{v_{1,2}2T_{s,d}%
}{\underline{t}_{1,2}},\quad d\Omega_{1}=\frac{A_{s}}{L_{1}^{2}},\quad
d\Omega_{2}=\frac{2dl}{L_{2}^{2}}%
\end{equation}
we have%

\begin{equation}
v_{1}\frac{\Sigma_{o}}{\left(  2\pi\right)  ^{3}}\frac{mv_{1}}{\hbar}%
\frac{v_{1}2T_{s}}{\underline{t}_{1}}k_{1}^{2}\frac{A_{s}}{L_{1}^{2}}\left(
2T_{o}\right)  \left(  2al\right)  \left(  \frac{1}{A_{s}}\right)  \frac
{d^{2}\sigma}{d\omega d\Omega}\frac{mv_{2}}{\hbar}\frac{v_{2}2T_{d}%
}{\underline{t}_{2}}\frac{2dl}{L_{2}^{2}}%
\end{equation}

for the number of scattered particles reaching the detector.

Collecting terms we have%

\begin{equation}
\frac{\Sigma_{o}}{\left(  2\pi\right)  ^{3}}\left(  \frac{m}{\hbar}\right)
^{4}\frac{v_{1}^{3}}{\underline{t}_{1}^{3}\underline{t}_{2}^{3}}%
l^{2}\mathbb{K}\frac{d^{2}\sigma}{d\omega d\Omega}%
\end{equation}
Comparing with (\ref{qint2}) we see that the two expressions agree if%

\begin{equation}
\frac{d^{2}\sigma}{d\omega d\Omega}=b^{2}\frac{v_{2}}{v_{1}}S\left(
\overrightarrow{q},\omega\right)
\end{equation}
which is equ. (26) of ref. \cite{vanh}.%

\newpage

\begin{center}
\textbf{Figure Captions}
\end{center}

Fig.1) Scattering in first Born approximation (equ. \ref{two}). The incoming
wave is scattered at $\overrightarrow{r}_{s}$ and the scattered wave
propagates according to $G_{o}$. The Green's function (\ref{greent}) has the
property that at a fixed time the wavelength decreases with distance from the source.

Fig. 2 a) The incident wave field, initially uncorrelated, emerges from the
entrance slit with correlations over a length scale $d$, inversely
proportional to the slit width, $a$.

\qquad2 b) Showing the phase relations responsible for the correlations of the
wave field according to the van Cittert-Zernike theorem. Waves emerging from
the uncorrelated point sources $a,b,c$ arrive at the scattering points
$r_{s},r_{s}^{\prime}$ with different relative phases. The phase difference of
the waves arriving at the different scattering points from each source point
depends on the location of the source point as well as the distance
$\overrightarrow{\delta}=\overrightarrow{r}_{s}-\overrightarrow{r^{\prime}%
}_{s}$, increasing with $\overrightarrow{\delta}$. This increasing phase
difference leads to a decrease in correlation between the waves arriving at
the points $r_{s},r_{s}^{\prime}$. Because the phase differences between the
uncorrelated source points are random they have no influence on the
correlation function.

Fig. 3) Illustrating equ. (\ref{biggy}). The functions $R_{in},R_{out}$
represent the effects of the phase differences between the paths 1 and 2 on
the correlation functions of the beam. $G_{s}$ is the correlation of the
density fluctuations at the different scattering points and $\left|  \psi
_{d}\right|  ^{2}$ is the sum of the contributions from paths going through
all possible pairs of points in the sample. Only pairs whose separation lies
within the correlation volumes contribute.

Fig. 4) Elastic scattering - general case. Showing the meaning of the symbols
used in the text. $\overrightarrow{\delta}$ is the distance between the two
scattering events at $\overrightarrow{r}_{s},\overrightarrow{r}_{s}^{\prime}$.
$\overrightarrow{\epsilon}_{s},\overrightarrow{\epsilon}_{s}^{\prime}$ are the
positions relative to center of the sample. Further symbols are defined in the text.

Fig. 5) Resolution function for small angle scattering

Fig. 6) Scattering from a time dependent system. The scatterer is assumed to
be concentrated at a single point $\left(  \overrightarrow{r}_{s}\right)  $.
The resolution is determined by the phase difference between paths (1) and
(2). Note the analogy with elastic scattering - fig. 4. $\tau$ is the delay
between the two scattering events at $t_{s},t_{s}^{\prime}$. $\xi_{s,}\xi
_{s}^{\prime}$ are the delays relative to the nominal scattering time
\underline{$t$}$_{s}$.


\newpage

\begin{thebibliography}{99}
\bibitem{vanh}L. van Hove, Phys. Rev., \textbf{95}, 249 (1954)

\bibitem {BW}M. Born and E. Wolf, \emph{Principles of Optics}, 6$^{th}$ ed.,
Pergamon Press, Oxford Sec. 10-4,

L. Mandel and E. Wolf, \emph{Optical coherence and quantum optics, }Cambridge
University Press, (1995)

\bibitem {ggtdo}R. G\"{a}hler, R. Golub, Z. Phys. \textbf{B56}, 5 (1984),

R. Golub, J. Felber, G. M\"{u}ller and R. G\"{a}hler, Physica \textbf{B162},
191 (1990)

\bibitem {nsplit}R. G\"{a}hler, R. Golub, K. Habicht, T. Keller and J. Felber,
Physica \textbf{B229,1 }(1996)

\bibitem {mezse}F. Mezei, ed., \emph{Neutron Spin Echo}, Proceedings of a
Laue-Langevin Institut Workshop, Oct. 1979. Springer-Verlag, 1980.

\bibitem {champ}D. C. Champeney, \emph{Fourier Transforms and their Physical
Applications}, Academic Press, N.Y., 1973

\bibitem {cz}P.H. van Cittert, Physica \textbf{1}, 201 (1934)

F. Zernike, Physica \textbf{5}, 785 (1938)

H.H. Hopkins, Proc. Roy. Soc. \textbf{A208}, 263 (1951)

\bibitem {pynnco}D. S. Sivia, R. N. Silver and R. Pynn, Nucl. Inst. Meth.
\textbf{A287}, 538 (1990)

\bibitem {ns}V. F. Sears, \emph{Neutron Optics}, Oxford, (1989), sec. 7.4

\bibitem {cohbab}R. Golub and S. K. Lamoreaux, Phys. Letts. \textbf{A162}, 122 (1992)

\bibitem {mezhouch}F. Mezei in J. P. Hansen, D.Levesque and J. Zinn-Justin,
\emph{eds.} Les Houches Session LI, July, 1989; \emph{Liquids, Freezing and
Glass Transition}, Elsevier, (1991)

\bibitem {synch}A.Q.R. Baron, A.I. Chumakov, H.F. Gr\"{u}nsteudel, H.
Gr\"{u}nsteudel, L. Niesen and R. R\"{u}ffer, Phys. Rev. Letts \textbf{77},
4308, (1996)

\bibitem {interlak}T. Keller, W. Besenb\"{o}ck, J. Felber, R. G\"{a}hler, R.
Golub, P. Hank and M. K\"{o}ppe, \emph{Proceedings of the Intern Conf on
Neutron Scattering}, Interlaken (1996), Physica \textbf{B234}, 1120 (1997).

J. Felber, R. G\"{a}hler, R. Golub, K. Prechtel, ''Coherence volumes and
neutron scattering'', \emph{to be published.}

\bibitem {marat}A.S. Marathay, \emph{Elements of Optical Coherence Theory,}
John Wiley, 1982

\bibitem {lips}B. A. Lippman and J. Schwinger, Phys. Rev. \textbf{79}, 469 (1950)

\bibitem {feynhib}R. P. Feynman and A. R. Hibbs, \emph{Quantum Mechanics and
Path Integrals}, McGraw-Hill , N.Y. (1965)

\bibitem {mosh}M. Moshinsky, Phys. Rev. \textbf{88},625 (1952)

\bibitem {mf}P. M. Morse and H. Feshbach, \emph{Methods of Theoretical
Physics}, Mcgraw-Hill, N.Y. (1953), Vol. 1, Chapter 7

\bibitem {SHULL}C. G. Shull and J. A. Oberteuffer, Phys. Rev. Letts.
\textbf{29}, 871 (1972)

J.~Arthur and M. A. Horne, Phys. Rev. \textbf{B32}, 5747 (1985)

J. Arthur,~C. G. Shull and A. Zeilinger, Phys. Rev. \textbf{B32}, 5753 (1985)
\end{thebibliography}
\end{document}